\documentclass[fleqn,usenatbib]{mnras}

\usepackage[T1]{fontenc}
\usepackage{wasysym}
\usepackage{txfonts}
\usepackage{xcolor}

\usepackage{lineno}

\DeclareRobustCommand{\VAN}[3]{#2}
\let\VANthebibliography\thebibliography
\def\thebibliography{\DeclareRobustCommand{\VAN}[3]{##3}\VANthebibliography}

\usepackage{graphicx, hyperref, url, lscape}	
\usepackage{xspace}
\usepackage{bm} 
\usepackage[position=top]{subfig}
\usepackage{multirow}

\def\prot{P_{\rm rot}}

\def\Rsun{R_{\star}}
\def\oA{\omega_{\rm A}}

\def\ta{t_{\rm A}}

\newcommand{\mean}[1]{\overline{#1}}

\title[Long-time evolution of Tayler Instability]
{Global simulations of Tayler instability in stellar interiors: a long-time multi-stage evolution of the magnetic field.
}

\author[Monteiro et al.]{
Monteiro, G.,$^{1,2}$\thanks{E-mail: guiaudias@ufmg.br}
Guerrero,  G.,$^{1,3}$
Del Sordo, F.,$^{4,5,6}$
Bonanno, A.,$^{6}$
Smolarkiewicz, P.K.$^{7}$
\\
$^{1}$Physics Department, Universidade Federal de Minas Gerais, Av. Antonio Carlos, 6627, Belo Horizonte, MG 31270-901, Brazil\\
$^{2}$Max-Planck-Institut für Sonnensystemforschung, Justus-von-Liebig-Weg 3, 37077 Göttingen, Germany\\
$^{3}$New Jersey Institute of Technology,  Newark, NJ 07103, USA\\
$^{4}$Institute of Space Sciences (ICE-CSIC), Campus UAB, Carrer de Can Magrans s/n, 08193, Barcelona, Spain\\
$^{5}$Institut d’Estudis Espacials de Catalunya (IEEC), 08034 Barcelona, Spain\\
$^{6}$INAF, Osservatorio Astrofisico di Catania, via Santa Sofia, 78 Catania, Italy\\
$^{7}$National Center for Atmospheric Research, Boulder, Colorado, USA}

\date{Accepted 2023 February 13. Received 2023 February 6; in original form 2022 November 17}

\pubyear{2023}

\begin{document}
\label{firstpage}
\pagerange{\pageref{firstpage}--\pageref{lastpage}}
\maketitle
\begin{abstract}
Magnetic fields are observed in massive Ap/Bp stars and are presumably present in the radiative zone of solar-like stars. To date, there is no clear understanding of the dynamics of the magnetic field in stably stratified layers. A purely toroidal magnetic field configuration is known to be unstable, developing mainly non-axisymmetric modes. Rotation and a poloidal field component may lead to stabilization. Here we perform global MHD simulations with the EULAG-MHD code to explore the evolution of a toroidal magnetic field located in a layer whose Brunt-Väisälä frequency resembles the lower solar tachocline. Our numerical experiments allow us to explore the initial unstable phase as well as the long-term evolution of such field. During the first Alfven cycles, we observe the development of the Tayler instability with the prominent longitudinal wavenumber, $m = 1$. Rotation decreases the growth rate of the instability and eventually suppresses it. However, after a stable phase, energy surges lead to the development of higher-order modes even for fast rotation. These modes extract energy from the initial toroidal field. Nevertheless, our results show that sufficiently fast rotation leads to a lower saturation energy of the unstable modes, resulting in a magnetic topology with only a small fraction of poloidal field, which remains steady for several hundreds of Alfven traveltimes. The system then becomes turbulent and the field is prone to turbulent diffusion. The final toroidal–poloidal configuration of the magnetic field may represent an important aspect of the field generation and evolution in stably stratified layers.

\end{abstract}




\section{Introduction}\label{sec:Introduction}
The mean-field dynamo theory \citep{Parker1955,SKR66} is the most comprehensible
framework to explain the origin of large-scale magnetic fields throughout the cosmos.
Comparison between observations, theoretical and numerical models allows 
for some constraining in the determination of the dynamo parameters.  
Nonetheless, explaining some cases of stellar magnetism with the mean-field theory is challenging.
Such is the situations of the $\sim 10 \%$ chemically
peculiar stars of types A and B, called Ap/Bp stars. 
These stars are characterized by magnetic fields
with simple tilted dipoles with amplitudes above 300 G \citep[see e.g., ][]{Donati2009ApBp,kochukhov+19} 
sustained in a radiative zone.  Concurrently, A-type stars like Vega
\citep{lignieres+09,petit+10} and Sirius A \citep{petit+11}
depict fields with less than 1 G, leaving a gap, also called {\it magnetic
desert}, between 1 and 300 G. The origin of this dichotomy is yet to be understood 
\citep[see for instance][]{Auriere2007,Szklarski+2013, Cantiello+19,Jermyn+20}, but any explanation certainly
relies on the evolution of magnetic fields in stably stratified layers (hereafter SSL). 
The dynamo mechanism faces also controversy in late type stars, like the
Sun,  having a convective envelope and a radiative interior. After the discovery
of the solar internal differential rotation \citep{SCHOU98}, the dynamo models have included the strong radial shear at the base of the convection zone as a main
constituent of the generation process 
\citep[see][for a review]{Ch20}. Nevertheless, observations
of stellar magnetic fields suggest that a general dynamo mechanism may exist 
for partially convective stars, where this shear layer is present as well as for fully convective stars \citep{Wright+16}. Global simulations 
have shown
that large-scale magnetic fields
can be generated in convection zones only, without the need for radial shear.
These simulations have obtained both steady \citep{BBBMT10} and oscillatory 
dynamo solutions \citep{Kapyla2012}. 
Both kind of solutions are also possible
in dynamo simulations including the tachocline \citep{Ghizaru2010,Guerrero2016}. 
Because in the latter simulations the field evolves in a region where the turbulent
diffusivity is reduced, the oscillatory solutions exhibit cycles with periods
closer to the solar one.  

\citet{LSC15, Guerrero2019} have proposed that magnetic instabilities
may play a relevant role in the field generation by inducing an $\alpha$-effect
in the radiative zone and ultimately defining the cycle period.  One class of 
dynamo operating in SSLs was heuristically described 
by \citet{Spruit2002}, revisited by \citet{Zahn2007} and implemented in a 
mean-field model by \cite{Bonanno13}. 
Thus, the evolution of magnetic fields
in radiative zones is instrumental for the understanding of the origin
of magnetic fields in early and late type stars. 

Thanks to the energy principle of \citet{Bernstein1958}, the spectral theory of static MHD plasma is rather well understood.
Using this principle \citet{Tayler1973}  demonstrated that a toroidal field,  $B_\phi$, can be unstable close to the axis of symmetry of a star. The stability conditions  
against axisymmetric and non-axisymmetric  perturbations
are $ d (B_\phi/s)/ds<0$, and  
$d (s B^2_\phi)/ds< 0$, where $s$ is the cylindrical radius, respectively.  
\citet{markey+73,markey+74} studied the stability of purely poloidal fields with
field lines closing inside the star. They found that this field configuration is also 
likely
unstable.  
In the case of cylindrical geometry, \citet{wright73} and \citet{tayler80} argued that mixed poloidal and toroidal field configurations
may be stable against adiabatic perturbations and could therefore exist inside  stellar interiors.  
This conclusion can  be violated in the presence of resonant modes with $\boldsymbol{B}\cdot\boldsymbol{k} = 0$, where $\boldsymbol{k}$ is the wavevector of the perturbation, making any mixed field configuration unstable to high longitudinal wave numbers, as discussed in \citet{Bonanno2012}.

For systems with rigid rotation, \citet{PittsTayler1985} 
found that rotation may help to stabilize the toroidal field. Complete stabilization, however,  can only be reached in the adiabatic case and considering unrealistic fast rotation. 
The effects of the Coriolis force opposing unstable perturbations may be different at equator and poles. Therefore, a latitudinal dependence is expected 
\citet{1980MNRAS.193..833G} 
(in spherical geometry this dependence may exists even in
the non-rotating situation because distance to the symmetry axis).  A general analytical study regarding the influence of rotation on the Tayler instability is complicated and has been restricted to simplified models.

The first MHD numerical experiments designed to address this problem where performed by \citet{Braithwaite2006StbTor}. They consisted of a local compressible model in Cartesian coordinates resembling a fraction of a star located at the north pole. The initial field was torus circulating along the axis of rotation. The time evolution of the magnetic field was restricted to the linear phase of the Tayler instability (TI) before reaching non-linear stages.  The results confirmed the theoretical findings. Particularly important for the present work is the verification that rotation stabilizes the magnetic field whenever the rotational frequency is larger than the Alfven frequency (AF), 
$\Omega_0 > \omega_{\rm A}$. Another important result is the confirmation that the Brunt-Väisälä frequency ($\omega_{BV}$), defined by the gravity acceleration and the
thermal stratification, imposes a lower limit for the radial wave number, i.e., the larger the $\omega_{BV}$, the smaller the wavelength of the unstable modes. This also exposes the dependency of the TI on the magnetic diffusivity, as magnetic fields with small spatial scales may diffuse before the instability develops \citep[see also][]{acheson80,Spruit1999}. \citet{Braithwaite2006StbTor} found that even at large values of magnetic diffusivity the instability is not fully suppressed but its growth rate decreases. Similarly, his analysis demonstrated that thermal
diffusion may diminish the effects of buoyancy allowing for larger wavelengths to be unstable for a given value of $\omega_{BV}$. Understanding the interplay between all these physical mechanisms is not straightforward. 

Another way of exploring the evolution of magnetic fields in SSLs is solving numerically the evolution of the linearized MHD equations for different Fourier modes. The behavior of a more realistic distribution of toroidal field in spherical geometry can be solved in rotating
systems in the purely adiabatic case or in the presence of dissipative terms.  \citet{KitchatinovRudiger2008} used this WKB approach in radius, while \citet{Bonanno2012,Bonanno2013} used it in latitude. These studies found that
the growth of the instability is dependent on the topology of the initial magnetic field. Also, \citet{Bonanno2012} found that the instability grows faster at the equator than at the poles, yet the most unstable mode is always $m=1$. According to their results, corresponding to adiabatic cases, solid body rotation may stabilize the magnetic field. 

Publications on global MHD simulations of this instability are scarce and have focused mainly in the evolution of an initial poloidal magnetic field in the presence of shear
\citep[e.g.,][]{Szklarski+2013,Jouve2015,Jouve2020}. For the most fundamental case of a toroidal field evolving in a SSL in spherical geometry, \citet{Guerrero2019Tayler} presented anelastic non-linear MHD simulations performed with the EULAG-MHD code. They explored the development of the Tayler instability of a toroidal field consisting of two bands of opposite polarity across the equator. The authors solved the inviscid MHD equations such that the dissipation is minimal
and delegated to the proven implicit-large-eddy simulation (ILES) property of the numerical algorithm \citep{Eulagmhd2013,Guerrero2022_EULAG_converg}. 
These global simulations were able to follow the evolution of the magnetic field beyond the linear phase. The authors explored the role of $\omega_{BV}$ in the stabilization of the initial magnetic field. Their results indicated that increasing $\omega_{BV}$, amounting to stronger buoyancy force, leads to growth rates that decrease following a power law, yet the instability is never fully suppressed.  Importantly, the number of radial modes also increases for large values of $\omega_{BV}$, as well as for large numerical resolution,  i.e., less dissipation, in agreement with the results of \citet{Braithwaite2006StbTor,KitchatinovRudiger2008}.
Thus, in simulations with higher resolution and large $\omega_{BV}$ the vertical extent of the unstable modes is so small that the instability becomes two-dimensional. 

This work is a continuation of \citet{Guerrero2019Tayler}.  However, in the models presented here the 
values of the Brunt-Väisälä frequency resembles the ones at the lower part of the solar tachocline.
Our main goal is exploring the stabilizing effect of solid body rotation, yet we also get insights on fundamental properties of the TI.  For instance, our setup allows us to verify the latitudinal dependence of the instability and the possible generation of helicity as a consequence of the growing of unstable modes \citep[see, e.g.,][]{stefani+19}.  In addition, our simulations 
allow us to explore the long term evolution and final morphology of the resulting magnetic field.

This paper is organized as follows. The description of the numerical model is presented in \S~\ref{sec:methods}. Our results and analysis for non-rotating and rotating cases are presented in \S~\ref{sec:results}. Finally, in \S~\ref{sec:conclusion}, we present the conclusions and discuss the implication of the results for the understanding
of magnetic fields in solar-like and Ap/Bp stars.  

\section{Numerical Simulations}
\label{sec:methods}

The numerical model solves the Lipps-Hemler anelastic system  of equations \citep{LippsHemler1982,Lipps1990} extended for the MHD case \citep{Eulagmhd2013}:
	\begin{equation}
	\boldsymbol{\nabla} \cdot \left(\rho_{\rm ad} \boldsymbol{u}\right) =0, \label{eq:dyn1}
	\end{equation}
	\begin{equation}
	\frac{D\boldsymbol{u}}{Dt}+2\boldsymbol{\Omega}\times\boldsymbol{u} = -\boldsymbol{\nabla}\left(\frac{p'}{\rho_{\rm ad}}\right)+\boldsymbol{g}\frac{\Theta^\prime}{\Theta_{\rm ad}}+\frac{1}{\mu_0\rho_{\rm ad}}\left(\boldsymbol{B}\cdot\boldsymbol{\nabla}\right)\boldsymbol{B},\label{eq:dyn2}
	\end{equation}
	\begin{equation}
	\frac{D\Theta^\prime}{Dt}=-\boldsymbol{u}\cdot\boldsymbol{\nabla}\Theta_{\rm amb}-\frac{\Theta^\prime}{\tau},\label{eq:dyn3}
	\end{equation}
	\begin{equation}
	\frac{D\boldsymbol{B}}{Dt}=\left(\boldsymbol{B}\cdot\boldsymbol{\nabla}\right)\boldsymbol{u}-\boldsymbol{B}\left(\boldsymbol{\nabla}\cdot\boldsymbol{u}\right).\label{eq:dyn4}
	\end{equation}
Here, $D/Dt=\partial/\partial t + \boldsymbol{u}\cdot\boldsymbol{\nabla}$
is the total time derivative, $\boldsymbol{u}$ is the velocity field
in a rotating frame with $\boldsymbol{\Omega} =\Omega_0(cos
\theta,-sin\theta, 0)$, where $\Omega_0 = 2\pi/T$ is the solid body
angular velocity and $T$ is the rotational period. The
pressure-perturbation variable, $p'$, accounts for both the gas and
magnetic pressure; i.e., $p^\prime = p^\prime_g +\frac{\boldsymbol{B}^2}{2\mu_0}$, with $\boldsymbol{B}$ and
$\mu_0$ denoting the magnetic field and permeability of free space.
The potential temperature perturbation, $\Theta^\prime$, enters both
the buoyancy term in the momentum Eq.~\ref{eq:dyn2} and the
entropy Eq.~\ref{eq:dyn3}; $\Theta$ is related to the specific
entropy via $s = c_p \ln\Theta + {\rm const}$, where $c_p$ is the
specific heat. The $\Theta^\prime$ perturbations are defined with
respect to a presumed ambient state, $\Theta_{\rm amb}$, typically
based on the stellar structure models. The last term on the rhs of
Eq.~\ref{eq:dyn3} relaxes $\Theta^\prime$ in a timescale $\tau = 5.184 \times
10^7\mathrm{s}$, unless stated otherwise  \citep[see][for a
substantive discussion]{Cossette2017}. The variables $\rho_{\rm ad}$ and
$\Theta_{\rm ad}$ are, respectively, the density and potential
temperature of the hydrostatic adiabatic reference state, whereas
$\boldsymbol{g} = g \hat{\bf e}_r$ is the gravity acceleration of
the solar interior adjusted from the solar model of \citet{CD+96}; hereafter JCD.
  
The model corresponds to a spherical shell with $0\leq \phi \leq 2\pi$, 
$0\leq\theta\leq\pi$, and ranging  from $r_b=0.6\Rsun$ to $r_t =0.76\Rsun$ in radius.  Eqs.~(\ref{eq:dyn1}-\ref{eq:dyn4}) are solved with the EULAG-MHD code \citep{Eulagmhd2013}
\footnote{The code is available at the dedicated website:
http://www.astro.umontreal.ca/~paulchar/grps/eulag-mhd.html}. 
The majority of the simulations are performed with a grid resolution of $126 \times 64 \times 28$ grid points in longitude $(\phi)$, latitude $(\theta)$ and radius $(r)$, respectively.
Parameters and results of this simulations are presented in Table~\ref{tab:results}, set A.  The equations are solved in their inviscid form, therefore, viscosity, thermal conduction and magnetic diffusivity are accounted by the truncation terms of the multidimensional positive-definite advection transport algorithm \citep[MPDATA;][]{smolarkiewicz2006}. Thus, the dissipation of the physical quantities changes when the resolution is increased or decreased. We evaluate how the grid resolution affects the results by performing simulations with double resolution in all directions  ($252 \times 128 \times 56$), or with double resolution in the radial direction only ($126 \times 64 \times 56$), Table~\ref{tab:results} sets B and C, respectively.

The boundary conditions are defined as follow: impermeable and stress-free at the top and bottom surfaces of the shell for the velocity field; radial field and perfect conductor
for the magnetic boundaries at the top and bottom boundaries, respectively;
the boundary conditions for $\Theta^\prime$ assume zero normal flux at the top and bottom boundaries.

\begin{figure}
  \centering
  \includegraphics[width=0.95\linewidth]{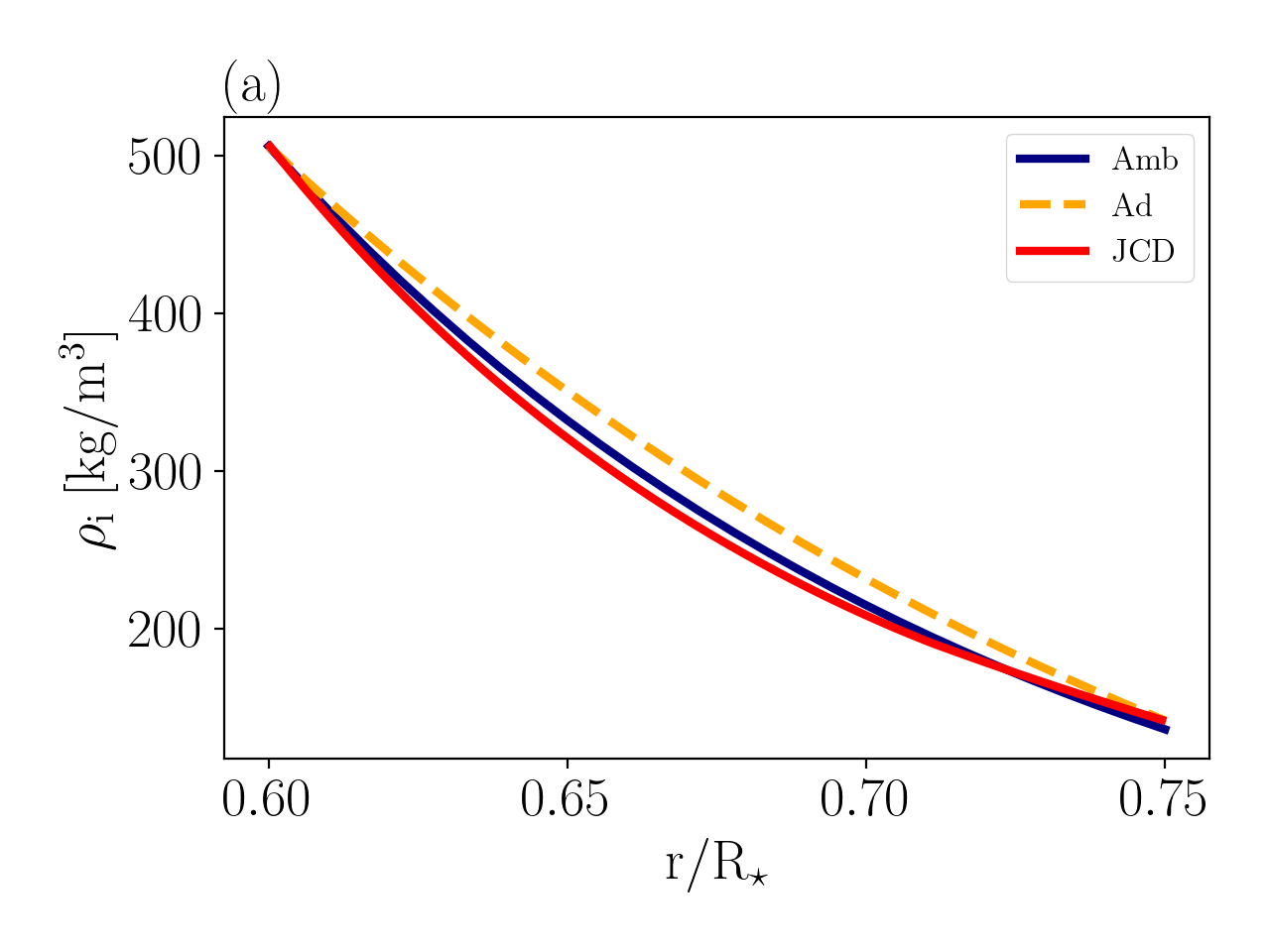}\\
  \includegraphics[width=0.95\linewidth]{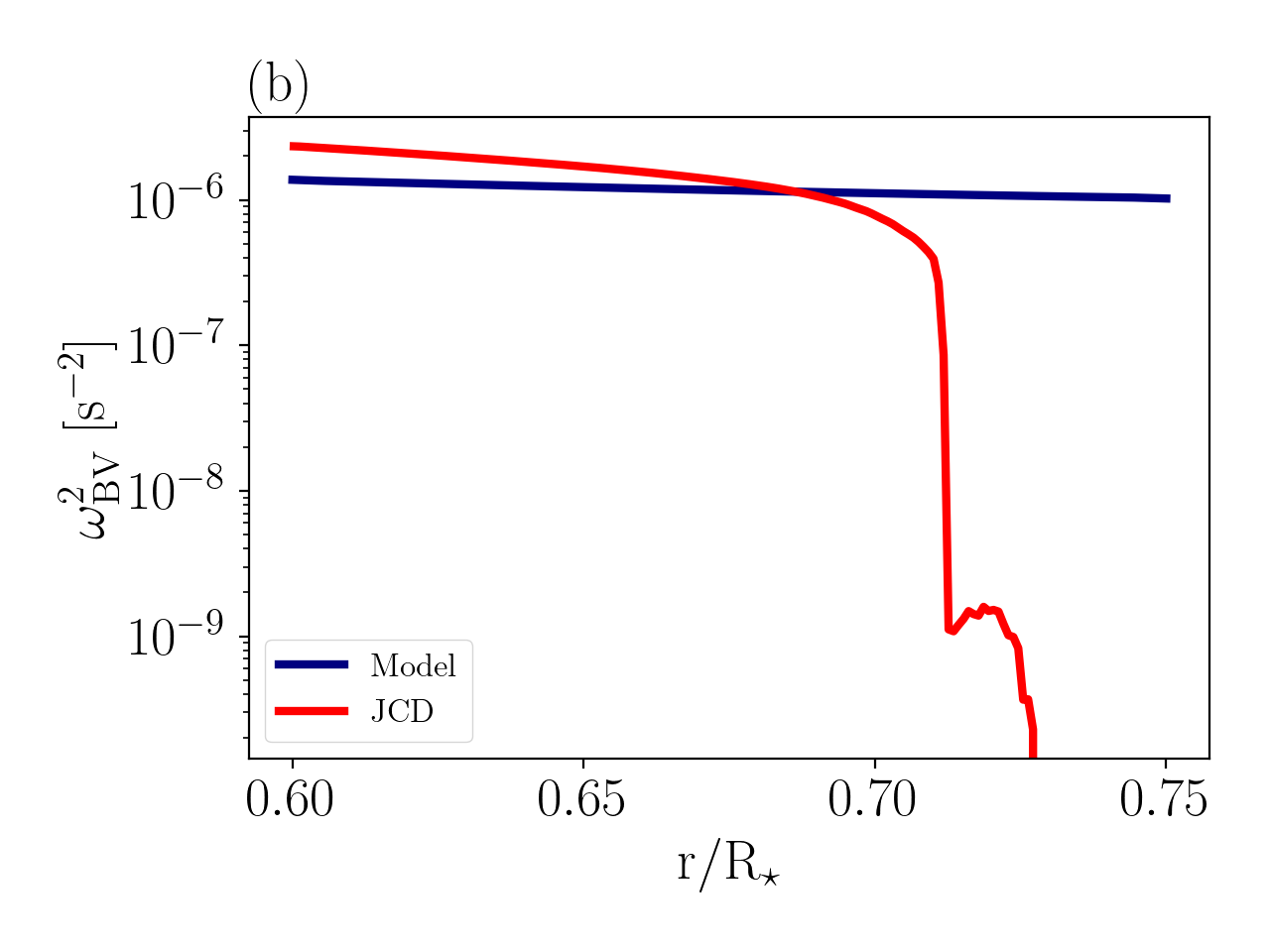}
  \caption{(a) Radial profile of ambient (blue line) and adiabatic (yellow dashed line) density stratification, and (b) square of the Brunt–Väisälä frequency for the ambient state (blue). For comparison the profiles corresponding to the JCD solar model are presented in red line. }
  \label{fig:structure}
\end{figure}

The ambient and adiabatic states are computed considering hydrostatic equilibrium in a non-rotating system, without magnetic field, for a polytropic gas. They are built by solving numerically the following equations for temperature, $T$, and density, $\rho$,
\begin{equation}\label{eq:temp_amb}
    \centering
    \frac{\partial T_i}{\partial r} = - \frac{g}{R_g\left(m_i+1\right)};
\end{equation}
\begin{equation}\label{eq:density_amb}
\centering
\frac{\partial\rho_i}{\partial r} = -\frac{\rho_i}{T_i}\left(\frac{g}{R_g}+\frac{\partial T_i}{\partial r}\right).
\end{equation}
The index $i$ stands for "ad" (adiabatic) or "amb" (ambient), and 
$R_g = 13732$ is the gas constant of a monatomic gas. 
Density, temperature and pressure are related via the equation of state for the perfect gas: $p_i = R_g\rho_iT_i$.
The boundary values at $r_b$, for solving Eqs.~(\ref{eq:temp_amb} and \ref{eq:density_amb}) are $T_b = 3.12 \times 10^6 \: \mathrm{K}$ and $\rho_b = 506 \: \mathrm{kg\: m^{-3}}$.  

In the simulations presented here we consider values of the Brunt-Väisälä frequency that are similar to those of the lower part of the solar tachocline.
Thus, the adiabatic and ambient profiles assume polytropic indexes $m_{ad}=1.5$  and $m_{amb} = 2.5$, respectively. Radial profiles of $\rho_{\rm amb}$ and $\rho_{\rm ad}$ are presented in 
Fig.~\ref{fig:structure}(a) with continuous blue and dashed yellow lines, respectively. The radial profile of the square of the Brunt–Väisälä frequency,
\begin{equation}
{\omega}_{B V}^{2}= \frac{g}{\Theta_{\mathrm{amb}}} \frac{\partial \Theta_{\mathrm{amb}}}{\partial r},\label{eq:BV_freq_1}
\end{equation}
is presented in Fig.~\ref{fig:structure}(b). For comparison, 
the red lines in both panels show profiles from the JCD model for the Sun.

\begin{figure}
	\begin{center}
	   \includegraphics[width=\columnwidth]{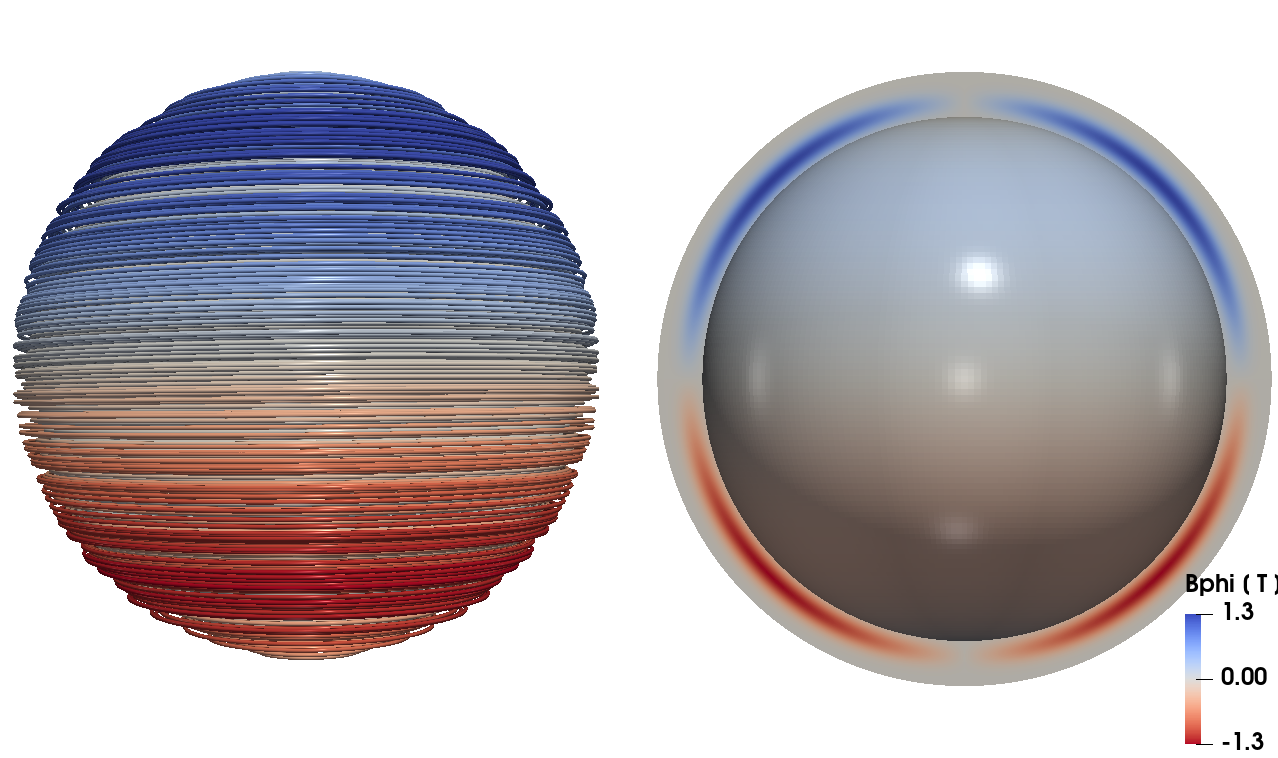}
	\end{center}
 \hfill	
\caption{
Configuration of the initial magnetic field (see Eq.~\ref{eq:antisym_magfield_ini}). The left panel presents the magnetic field lines at $r=0.68 \Rsun$.  The right panel shows the field distribution in the ($\phi-\theta$) plane at $r=0.64 \Rsun$ and in the meridional ($r,\theta$) plane.  Red (blue) colors correspond to toroidal field circulating clockwise (counterclockwise) with magnitudes presented in the color bar.}
\label{fig:initial_B}
\end{figure}

The initial magnetic configuration is a toroidal field anti-symmetric across the equator, i.e.,  $B_{r0}=B_{\theta 0} = 0$ and
\begin{eqnarray}
B_{\varphi 0}(r,\theta,\phi)  &=& B_0f(r)\sin \theta \cos\theta \hat{e}_{\varphi}, 
\label{eq:antisym_magfield_ini}
\end{eqnarray}
with
\begin{equation}
f(r) = \exp\left[-\frac{(r-r_c)^2}{d^2} \right],
\label{eq:radial_prof}
\end{equation}
where $r_c=0.675\Rsun$, $d=0.03\Rsun$. This configuration is used in all simulations presented in this paper. $B_0$ is the magnitude of the initial magnetic field and is a free parameter of the model.  The field lines distribution corresponding to this magnetic field is presented in the left panel of Fig.~\ref{fig:initial_B} (left panel). The profile of the magnetic field in the meridional plane is shown in the right panel of Fig.~\ref{fig:initial_B}.
The red and blue colors in the figure represent clockwise and counterclockwise orientation of the toroidal field. Notice that the radial profile considered here is different from the one used in previous studies \citep[e.g.,][]{Braithwaite2006StbTor, KitchatinovRudiger2008,Bonanno2012}. This initial field is a reasonable representation of a toroidal field resulting after the winding-up of a poloidal field due to differential rotation.      For the velocity field and the thermal perturbations the initial conditions are zero.

\section{Results}
\label{sec:results}
\begin{table*}
    \centering
    \resizebox{1.5\columnwidth}{!}{\begin{tabular}{clcccccccc}
       \multicolumn{1}{c}{Set}&
       \multicolumn{1}{c}{Model} &  \normalsize $\Omega_0\cdot 10^{-6}$ & $\overline{\omega}_A\cdot10^{-8}$ & $\overline{\omega}_{BV}\cdot10^{-3}$ & $\delta\cdot10^4$ &$\eta$ & $B_P/B_T$ & $\sigma \cdot 10^{-1}$ $[\rm \omega^{-1}_A]$ & $|\Gamma|\cdot 10^{-9}$  \\
    \hline
    \hline
        \multirow{11}{*}{\rotatebox[origin=c]{90}{\parbox[c]{0.5cm}{\centering A1}}}
           &A\_NRB10  	    & 0     & 7.35 & 1.08 & 1.48 &   0    & 0.11 & 0.69 &6.25 \\
           &A\_R300B10  	& 0.24  & 7.35 & 1.08 & 1.48 & 6.70   & 0.09 & 0.41 &3.31 \\    
    	&A\_R150B10  	& 0.49  & 7.35 & 1.08 & 1.48 & 13.40  & 0.04 & 0.38 &2.47 \\
    	&A\_R130B10  	& 0.56  & 7.35 & 1.08 & 1.48 & 15.46  & 0.05 & 0.40 &2.57 \\
    	&A\_R110B10  	& 0.66  & 7.35 & 1.08 & 1.48 & 18.26  & 0.08 & 0.41 &2.49 \\
    	&A\_R70B10  	& 1.04  & 7.35 & 1.08 & 1.48 & 28.71  & 0.08 & 0.27 &2.23 \\
    	&A\_R50B10  	& 1.45  & 7.35 & 1.08 & 1.48 & 40.19  & 0.05 & 0.28 &2.26 \\
    	&A\_R30B10  	& 2.42  & 7.35 & 1.08 & 1.48 & 66.98  & 0.08 & 0.22 &1.94 \\
	    &A\_R20B10  	& 3.64  & 7.35 & 1.08 & 1.48 & 98.94  & 0.07 & 0.16 &1.98 \\
	    &A\_R10B10  	& 7.27  & 7.35 & 1.08 & 1.48 & 197.80 & 0.06 & 0.04 &1.55 \\
    \hline
        \multirow{3}{*}{\rotatebox[origin=c]{90}{\parbox[c]{0.5cm}{\centering A2}}}
        &A\_NRB10$\tau05$ 	& 0     & 7.35 & 1.08 & 1.48 &   0        & 3.29 & 3.28 &7.19 \\   
        &A\_NRB10$\tau5 $ 	& 0     & 7.35 & 1.08 & 1.48 &   0        & 1.44 & 1.46 &7.26 \\ 
        &A\_NRB10$\tau50$ 	& 0     & 7.35 & 1.08 & 1.48 &   0        & 0.74 & 0.71 &6.94 \\ 
    \hline
        \multirow{4}{*}{\rotatebox[origin=c]{90}{\parbox[c]{0.5cm}{\centering A3}}}
        &A\_R10B80  	& 7.27  & 58.81& 1.08 & 0.18 & 24.73   & 0.03 & 0.30 & 9.88 \\
        &A\_R10B40  	& 7.27  & 29.4 & 1.08 & 0.37 & 49.45   & 0.06 & 0.47 & 5.95 \\
        &A\_R10B20  	& 7.27  & 14.7 & 1.08 & 0.74 & 98.90   & 0.06 & 0.113 & 3.56\\
        &A\_R10B5  	& 7.27  & 3.68 & 1.08 & 2.95 & 395.76      & 0.06 & 0.39 & 0.56 \\         
   \hline
        \multirow{3}{*}{\rotatebox[origin=c]{90}{\parbox[c]{0.5cm}{\centering B}}}
        &B\_NRB10    &  0    & 7.42 & 1.08 & 1.46 &   0    & 0.09 & 0.36 &3.16 \\
    	&B\_R300B10  & 0.24  & 7.42 & 1.08 & 1.46 & 6.70   & 0.17 & 0.40 &1.93 \\  
        &B\_R10B10   & 7.27  & 7.42 & 1.08 & 1.46 & 197.80 & 0.05 & 0.08 & 0.87 \\    	
    \hline
        \multirow{2}{*}{\rotatebox[origin=c]{90}{\parbox[c]{0.5cm}{\centering C}}}
    	&C\_R300B10  & 0.24   & 7.35 & 1.08 & 1.48 & 6.70   & 0.07 & 0.46 & 2.83  \\
   	    &C\_R10B10   & 7.27   & 7.35 & 1.08 & 1.48 & 197.80 & 0.03 & 0.06 & 1.56\\

    \hline
    \end{tabular}}
\caption{Parameters and results of the simulations. The rotational rate is 
given by $\Omega_0$, with $\overline{\omega}_{BV}$,  $\delta$ and  $\eta$ defined in Eqs.~{\ref{eq:BV_freq}, \ref{eq:deltaeta}}, respectively. 
$B_P/B_T$ shows the ratio between poloidal and toroidal fields at the end of 
linear phase. $\sigma$ is the growth of $m=1$ magnetic longitudinal mode on early Alfven Cycles, while
$|\Gamma|$ represents the calculated decrease rate of the $m=0$ longitudinal mode.
Sets are separated by resolution, field intensity and/or relaxation time. Meanwhile the simulation acronym specifies set, rotation period in earth days and field intensity.
}
    \label{tab:results}
\end{table*}

The models presented here are similar to the ones studied by \citet{Guerrero2019Tayler}, except that in this paper the radial extent of the shell is thinner, with 
values of Brunt-Väisälä frequency that resembles the lower part of solar tachocline.
The simulations are characterized by the non-dimensional parameters  
\begin{equation}
\delta^2 = \frac{\mean{\omega}^2_{B V}}{\mean{\omega}^2_A}, \quad \; {\rm and} \; \quad
\eta^2 = \frac{4\Omega^2_0}{\mean{\omega}^2_A},
\label{eq:deltaeta}
\end{equation}
where 
\begin{equation}
\mean{\omega}^2_{B V}=\left\langle\frac{g}{\Theta_{\mathrm{amb}}} \frac{\partial \Theta_{\mathrm{amb}}}{\partial r}\right\rangle_{r} ,\label{eq:BV_freq}
\end{equation}
and
\begin{equation}
    \mean{\omega}^2_A =\left\langle\frac{B^2_{\phi 0}}{\mu_0 \rho_{amb} \varpi^2_c}\right\rangle_{r,\theta}, \label{eq:Alfven_freq}
\end{equation}
are averages of the Brunt-Väisälä and Alfven frequencies, in radius, and radius and latitude, respectively. The length scale considered in the Alfven frequency is the level arm at $45^{\circ}$, $\varpi_c = r_c\sin(\pi/4)$.  
The time units presented in this work are normalized with $\ta = 1/\overline{\omega}_A$. This quantity represents the number of cycles that an Alfven wave travels along toroidal field lines with radius $\varpi_c$.

\subsection{Non-rotating cases} 
\label{subsec:results_nr}

Ideally, simulations should start from a state in magnetohydrostatic equilibrium i.e., the gas pressure should balance the magnetic pressure. This basic state would then be perturbed to drive the onset of the instability. This was done, for instance, by \citet{BBDSM12} to study the helical symmetry breaking occurring during the TI. However, this study uses a more complex setup than that of \citet{BBDSM12}, and finding this equilibrium state is difficult, especially in the rotating cases where the Coriolis force is also present.  In this way, the hydrostatic equilibrium of the ambient state, Eqs.(\ref{eq:temp_amb} and \ref{eq:density_amb}), is unbalanced by the initial magnetic field.  
After a few time steps, a new balance is achieved with the development of radial and latitudinal profiles of $\Theta^\prime$ and ${\bf u}$. These profiles have most of their energy in mode $m=0$, yet, due to numerical noise (i.e. round-off errors)  other modes with $m>0$ are also existent through the domain. Although their energy is compatible with numerical precision, they are sufficient to trigger the TI (a view on this process can be observed in the movies {\tt A\_R10B10.mp4} and {\tt B\_R10B10.mp4} submitted as complementary material).

\begin{figure}
  \includegraphics[width=\linewidth]{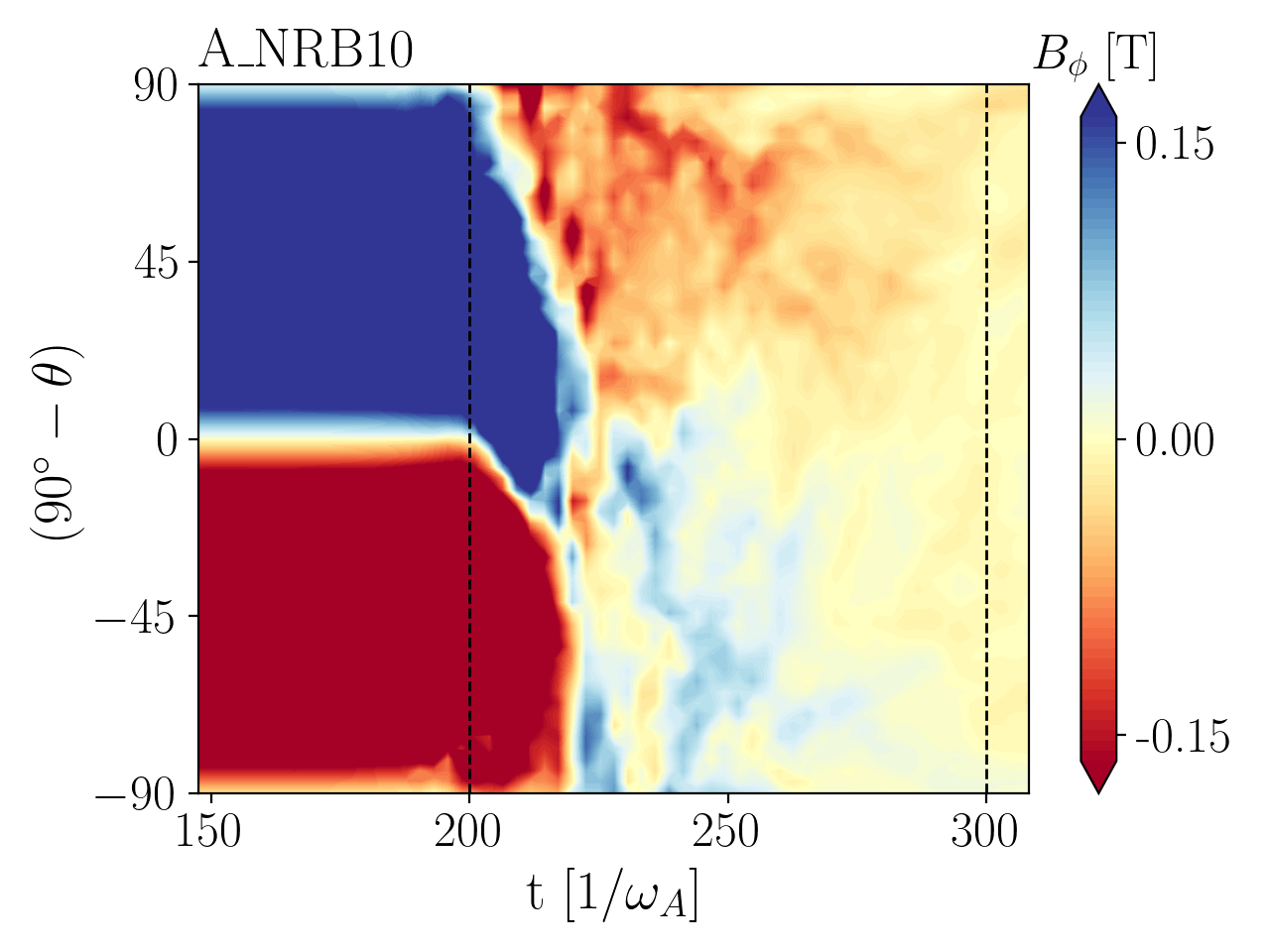}
  \caption{Time-latitude diagram presenting the evolution of $B_{\phi}$ for the non-rotating case A\_NRB10. The field corresponds to $r=0.7 \Rsun$
  and $\phi=90^{\circ}$. The red (blue) colors represents clockwise (counter clockwise) toroidal magnetic field.  Time units are expressed in Alfven travel times.}\label{fig:Anti_nr_10b_longt_ev}
\end{figure}

\begin{figure}
  \includegraphics[width=\columnwidth]{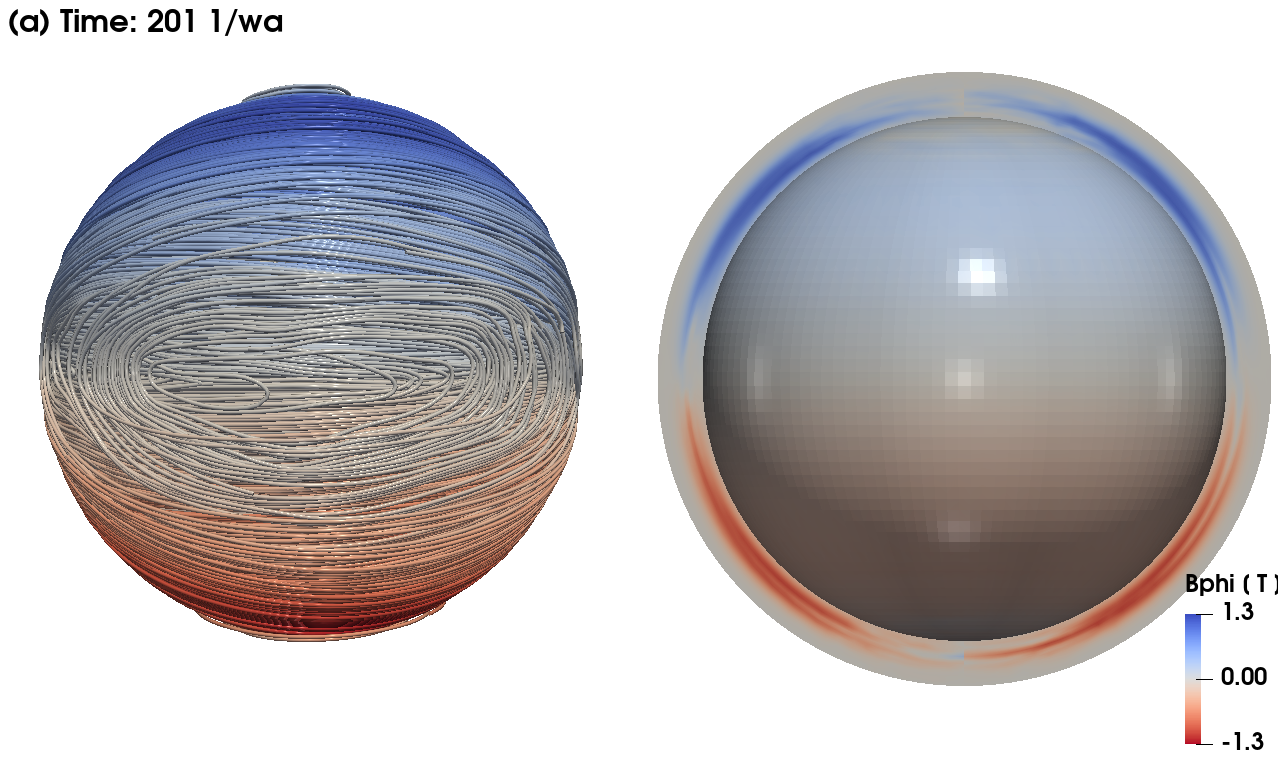}
  \includegraphics[width=\columnwidth]{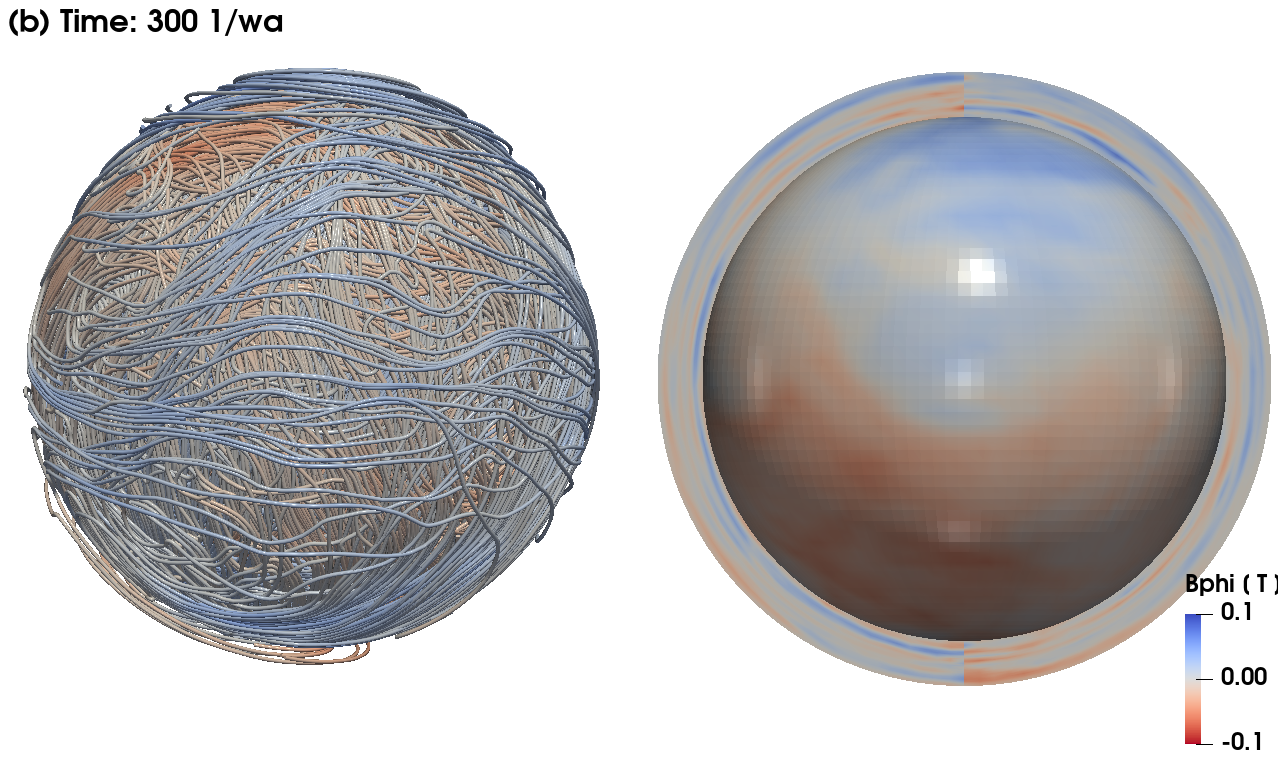}
\caption{Similar to Fig.~\ref{fig:initial_B} 
for the simulation A\_NRB10 at time marks (a) $201$ and (b) $300$ ($1/\omega_A$). These times are marked by black dashed lines in Fig.~\ref{fig:Anti_nr_10b_longt_ev}.}
\label{fig:Anti_nr_10b_snap}
\end{figure}

Fig.~\ref{fig:Anti_nr_10b_longt_ev} shows the evolution of the toroidal magnetic field in latitude and time, at $r = 0.7 \Rsun$ and $\phi = 90^\circ$.
A noticeable decay of the initial field occurs after the unstable modes reach their maximum energy value. The decay follows a wave-like pattern, with minor changes for different longitudes (not shown). 
After the field decays it rapidly dissipates.
These pattern resembles the results obtained by \citet{Guerrero2019Tayler} (e.g., see their Fig. 7). 
The configuration of the magnetic field lines at the instant  when $B_{\phi}$ experiences a significant decay is presented in Fig.~\ref{fig:Anti_nr_10b_snap}(a). The figure shows the tilt of the magnetic lines with respect to the symmetry axis, and the consequential displacement of magnetic axis at polar regions and the opening of the field lines at the equator. These are all manifestations of the hydromagnetic instability.  Fig.~\ref{fig:Anti_nr_10b_snap}(b) 
shows the field lines after at $300$ Alfven travel times. At this moment, the magnitude of the field has markedly decreased and the magnetic topology is substantially different from the initial one.
The black dashed lines in Fig.~\ref{fig:Anti_nr_10b_longt_ev}
indicate the times where these snapshots were taken.
Note that similar effects of the instability occur at the equator and the poles. Although they are not simultaneous (see below), the growth rate at both places is similar.  

The study of the time evolution of different longitudinal modes at different latitudes can
be performed by computing the power spectrum of the magnetic and/or the kinetic energy
integrated over latitudinal bands at one specific radius. The magnetic energy as function of the spectral index $m$ and time is defined as,
\begin{equation}
\tilde{E}_B(m,t) = \tilde{B^R_{\phi}}(m,t)^2  + \tilde{B_{\theta}}(m,t)^2 + \tilde{B_{r}}(m,t)^2,
\label{eq:power_spec}
\end{equation}
where the tildes represent the quantities in the Fourier space. For example, 
$\tilde{B^R_{\phi}}$ is computed as 
\begin{equation}
\tilde{B}_{\phi}^R = \frac{1}{2 \pi}  \frac{1}{(\theta_2-\theta_1)}  \int_0^{2\pi} \int_{\theta_1}^{\theta_2} B_{\phi}^R \exp (-i m \phi) d \theta d \phi.
\label{eq:fourrier_trans}
\end{equation}
Note that $B_{\phi}^R$ is the residual,
\begin{equation}
B_{\phi}^R(r_c,\theta,\phi,t) = B_{\phi 0}(r_c,\theta,\phi,t) - B_{\phi} (r_c,\theta,\phi, t).
\end{equation}
This subtraction is only needed for the toroidal component.  Three different latitudinal intervals were considered, $[\theta_1, \theta_2]$, defined as 
$NP=[0,\pi/9]$,  $EQ=[\pi/3, 2\pi/3]$, and $SP = [8\pi/9, \pi]$. These are the latitudes where changes in the field topology are initially observed.  Similar equations are used to compute the spectral kinetic energy. Throughout the paper references to spectral energies corresponds to quantities evaluated using these transformations.  For the TI, the development of all the modes is expected, yet $m=1$ is predicted to be the fastest growing one \citep{Goedbloed2010book}.

\begin{figure}
  \includegraphics[width=\columnwidth]{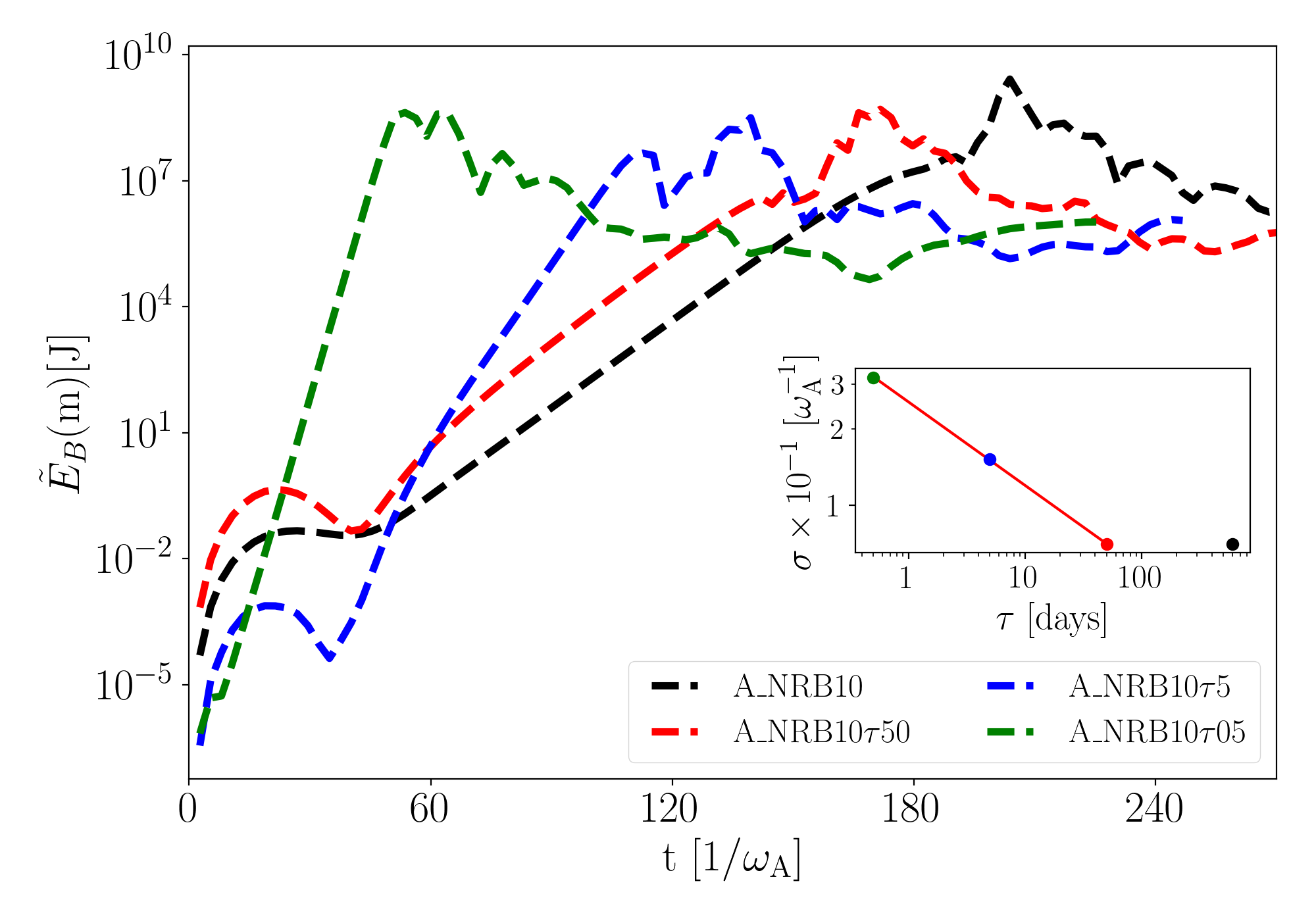}
\caption{
Temporal evolution of the $m=1$ mode at the south pole for the non-rotating cases from set A2. Different colors correspond to values of $\tau$ of 600 days (black, set A1), 50 days (red), 5 days (blue) and 0.5 days (green). 
}
\label{fig:nr_m1_comparison}
\end{figure}

As was demonstrated analytically by \citet{Zahn1974}, the resistance of the plasma in stellar interiors to dynamical instabilities depends strongly on the thermal timescale and/or on the BV frequency. In the context of the TI, such results have been confirmed by \citet{Braithwaite2006StbTor} for changes in BV frequency and the thermal conductivity, and by \citet{Guerrero2019Tayler} for changes in BV frequency only. The evolution of unstable parcels of the fluid depends on the thermal properties of the gas, represented in the prognostic equations by $\Theta^\prime$. In these equations, the thermal timescale is defined through the relaxation time, $\tau$ (Eq.~\ref{eq:dyn3}). Smaller values of $\tau$ result in small amplitudes of short lived thermal perturbations, and vise-versa. Thus, it is possible to explore the thermal effects on the TI by changing $\tau$. 
The set of simulations A2, together with simulation A\_NRB10, encompass values of $\tau$ between $0.5$ and $600$ days. 
The temporal evolution of the mode $m=1$ for these simulations is presented in Fig.~\ref{fig:nr_m1_comparison}. The figure shows that the growth rate, $\sigma$, decreases with the increase of $\tau$ following a power law, $\sigma \propto \tau^{-0,32}$ (see red line in the inset panel). The response of the plasma to changes in $\tau$ is similar to the response when increasing the BV frequency \citep{Guerrero2019Tayler}. In both situations the TI is not fully suppressed but grows slowly. Note also that for sufficiently large values of $\tau$, there is no change in $\sigma$. This happens because the amplitude of $\Theta^\prime$ depends on the smaller timescale in the model. Thus, for a long relaxation timescale it will be independent of $\tau$ and rely on the Alfvenic timescale ($\sim 150$ days for the value of $B_0$ used in this set of simulations).

\begin{figure*}
\includegraphics[width=\linewidth]{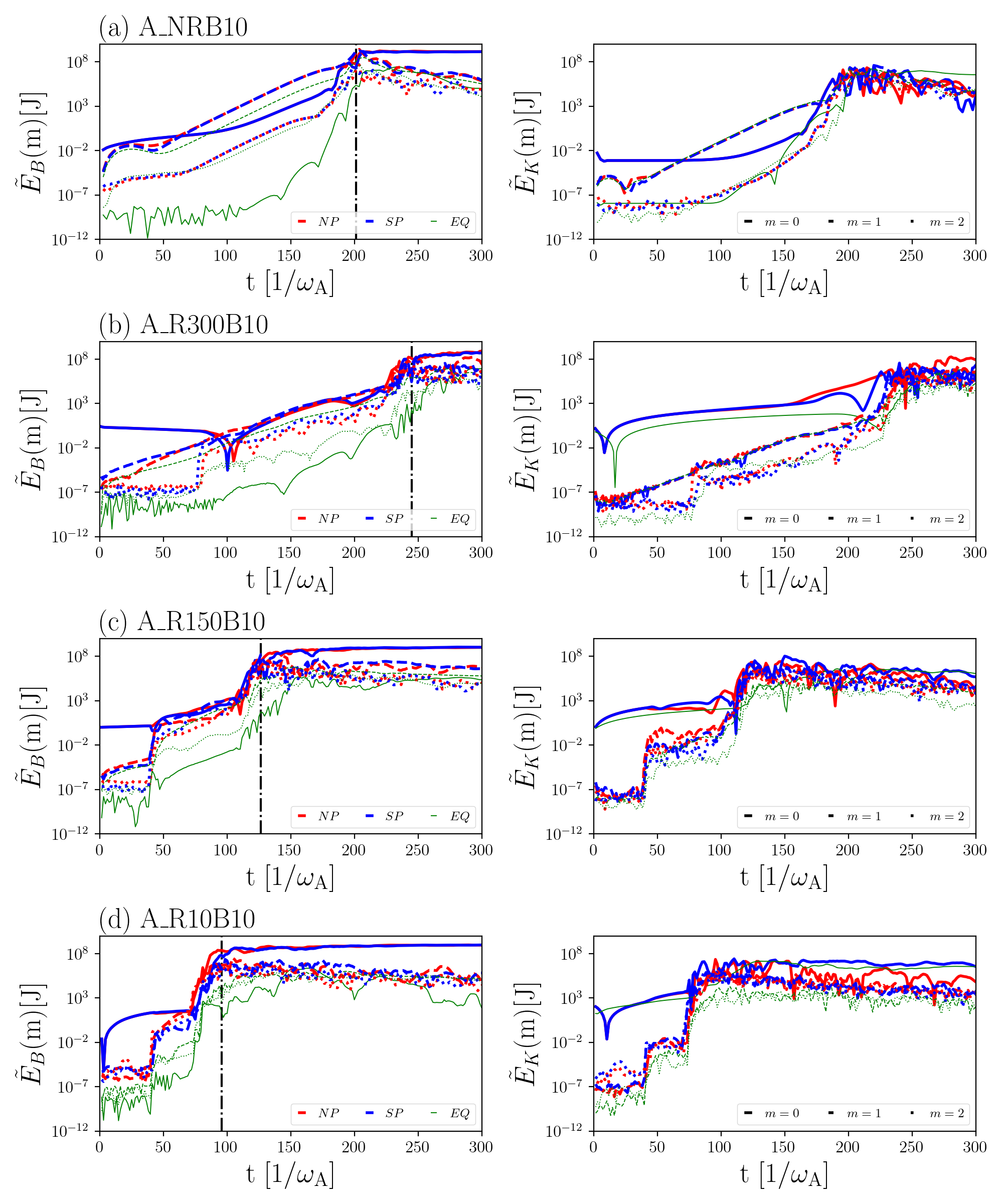}
\caption{ Evolution of the modes $m=0$ (Eq.~\ref{eq:fourrier_trans}, continuous), $m=1$ (dashed), and $m=2$ (dotted) of the magnetic (left panels) and kinetic (right panels) energies. Panels (a) to (d) correspond to simulations A\_NRB10 (no rotation), A\_R300B10 ($\prot = 300$ days), A\_R150B10 ($\prot = 150$ days), and A\_R10B10 ($\prot = 10$ days), respectively. Different colors correspond to the latitudes where the energy of the unstable modes is computed, the red and blue lines correspond to the north and south poles, and the thin green lines correspond to the equator.  The black dot-dashed lines indicate the saturation phase (see the text).}
  \label{fig:m_energy_evo_std}
\end{figure*}

In Fig.~\ref{fig:m_energy_evo_std}(a) we present the time evolution of the modes $m=0$ (continuous lines),  $m=1$ (dashed lines) and $m=2$ (dotted lines) of the magnetic (left) and kinetic (right) spectral energy density for the non-rotating simulation A\_NRB10. The colors stand for polar regions $NP$ (red), $SP$ (blue), and $EQ$ (thin green lines). 
At the beginning of evolution, after a transient phase 
where the fluid balances the magnetic forces, the energy of the longitudinal modes rises. The modes $m=1$ and $m=2$ grow faster than $m=0$. Similar behavior is observed in the kinetic energy spectra. The instability develops earlier at polar regions, nevertheless, there is no significant difference in the growth rate between pole and equator for modes $m=1$ and $2$.  The mode $m=0$ develops faster at the poles.  The growth of the mode  $m=0$ alongside $m=1$ is an expected result for a steeper radial profile of the magnetic field \citep{Braithwaite2006StbTor}.

The magnetic field topology is substantially affected by the development of unstable modes only when the energy of these modes is comparable to the initial magnetic energy.  The deformation of the field lines at this stage is displayed in Fig~\ref{fig:Anti_nr_10b_snap} (a). After further evolution, the field lines loss any resemblance with the initial configuration, see Fig~\ref{fig:Anti_nr_10b_snap} (b).

From the previous qualitative and quantitative analysis, three different stages of evolution are identified:
(a) the linear phase,  where the unstable modes grow exponentially;  (b) the saturation phase, where the energy of these modes reaches similar values than the initial magnetic energy;  (c) the decaying phase where the magnetic field diffuses.  As it will be shown later, this phase is compatible to turbulent decay following the Kolmogorov scaling, $\hat{E}_B \propto m^{-5/3}$.  

In the following sections we will study the behavior of the magnetic and velocity fields in these three phases for simulations with various rotation rates. We will also explore the effects of different resolution on the magnetic field evolution during these phases.

\subsection{The stabilizing effect of rotation}
\label{results:rotation}

\begin{figure}
\includegraphics[width=\columnwidth]{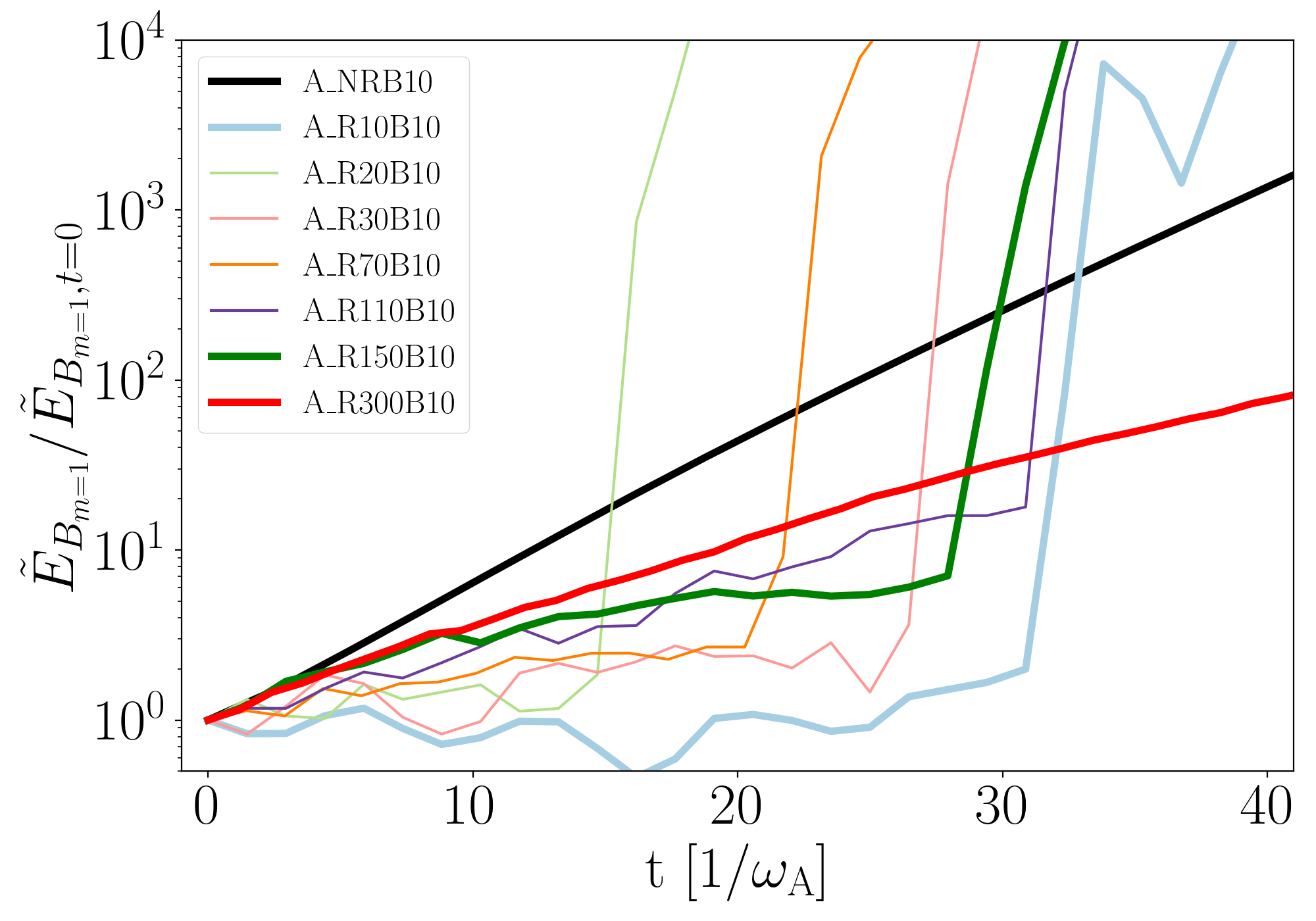}
\caption{
Early temporal evolution of magnetic mode $m=1$ for the same simulations displayed in Fig.~\ref{fig:m_energy_evo_std}. For a better comparison of the growth rates the energy is normalized by its value at $t=0$, and the initial transient phase 
has been removed. 
}
 \label{fig:m1_energy_comp_begin}
\end{figure}

\begin{figure}
 \includegraphics[width=\columnwidth]{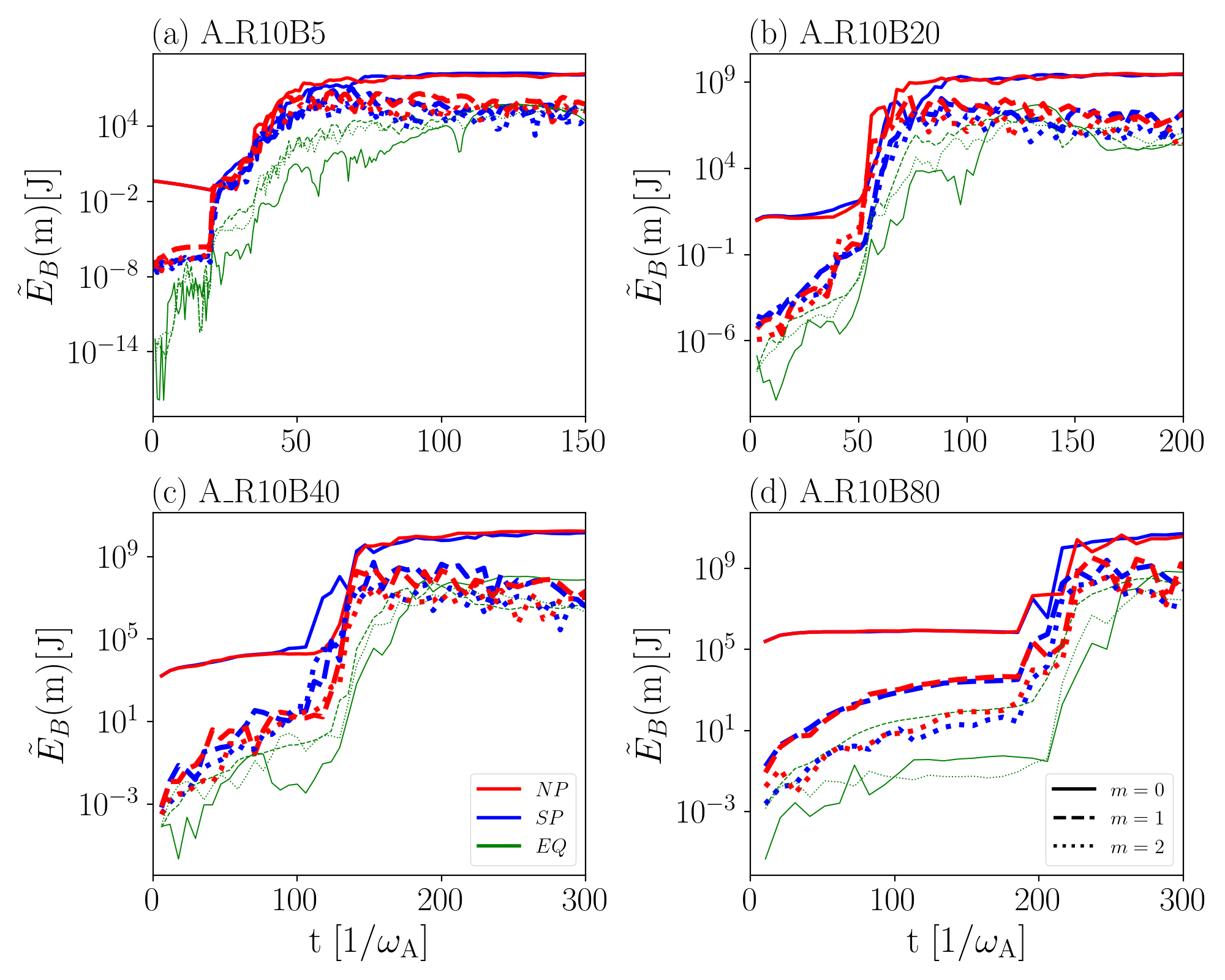}
\caption{Same as Fig.~\ref{fig:m_energy_evo_std} for simulations with different magnitude of initial magnetic field (a) A\_R10B5 ($B_0 =0.65$ T), (b) A\_R10B20 ($B_0 =2.6$ T), (c) A\_R10B40 ($B_0 =5.2$ T), and (d) A\_R10B80 ($B_0 =10.4$ T).}
\label{fig:r10_m1_comparison_wa}
\end{figure}

\begin{figure*}
\includegraphics[width=\linewidth]{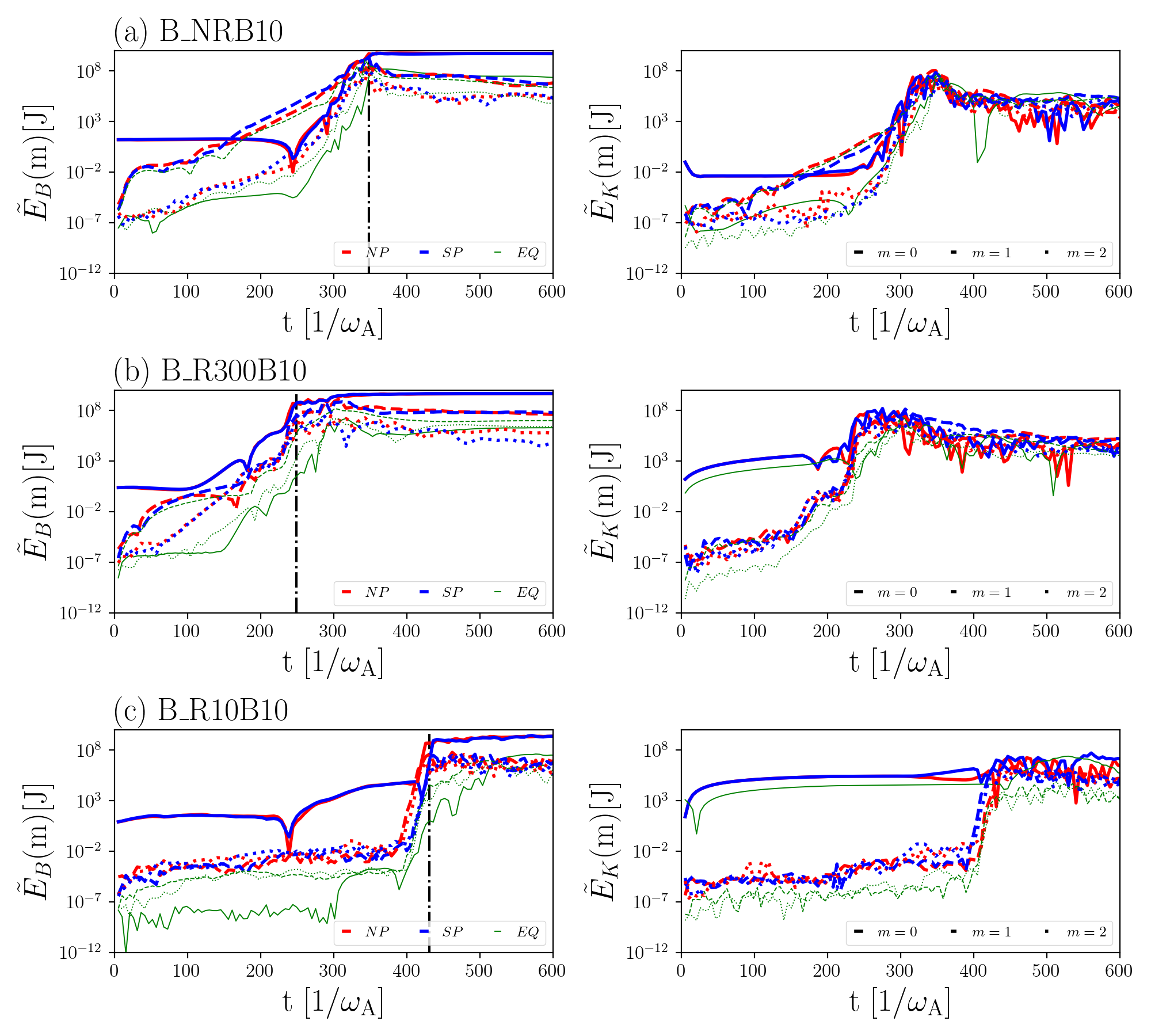}
\caption{Evolution of the energy of initial modes of the doubled resolution simulations. Line style and color properties are the same as the ones used on Fig.~\ref{fig:m_energy_evo_std}.}

 \label{fig:m_energy_evo_high}
\end{figure*}

Several authors have shown that rotation helps stabilizing toroidal fields in radiative zones whenever $\Omega_0$ is comparable to $\overline{\omega_{\rm A}}$ ($\eta \sim 1$) \citep[e.g.,][]{Braithwaite2006StbTor,Bonanno2013}. 
Nevertheless, if the system has any finite diffusion, full stabilization will only be achieved if $\omega_{\rm A} \ll \Omega_0$ ($\eta \gg 1$) \citep{KitchatinovRudiger2008}.  
Otherwise, rotation will only decrease the growth rate of the instability. All these works have focused on the linear phase but have not explored the saturation nor the decaying phase of the field.

The simulations presented in this section include the Coriolis term in Eq.~\ref{eq:dyn2} to explore if the TI can be stabilized by rotation.  In the set of simulations A1,  $\omega_A$ is kept constant and the rotation rate, $\Omega_0$ varies between $0.24 \times 10^{-6} \; {\rm Hz}$ and $7.27 \times 10^{-6} \; {\rm Hz}$, corresponding to rotation periods between 300 and 10 days, and $6.7 \le \eta \le 197.8$.  Alternatively, in the set of simulations A3 the rotation is kept constant, $\Omega_0 = 7.27 \times 10^{-6} \; {\rm Hz}$ (10 days), and $\omega_A$ varies in the range $3.68 \times 10^{-8} \; {\rm Hz} \leq \oA \leq 58.81 \times 10^{-8} \; {\rm Hz}$.  This corresponds to values of $B_0$ between $0.65$ and $10.4$ T, and $24.7 \le \eta \le 395.7$.  The simulations evolve for several hundreds of Alfven travel times to consider the three phases of evolution described above.  

\subsubsection{Linear phase}

Exploring the effects of changing $\Omega_0$ and $\omega_A$ in the linear phase of the TI allows for comparison between global non-linear simulations and the earlier theoretical predictions \citep[see e.g.][]{PittsTayler1985} and numerical results \citep[see e.g.][]{Braithwaite2006StbTor}. Figure~\ref{fig:m_energy_evo_std} (panels b to d) shows the time evolution the magnetic (left) and kinetic (right) energy densities in simulations A\_R300B10, A\_R150B10 and A\_R10B10, respectively. 
The effects of including rotation in the simulation A\_R300B10 ($\eta\sim 7$) can be observed in Fig.~\ref{fig:m_energy_evo_std}(b). The linear evolution of mode $m=1$ in the log-normal plot is an important indication that this value of $\eta$ is not sufficient to suppress the TI. It is interesting to notice that no significant difference are observed between lower (see green dashed lines) and high latitudes (red and blue dashed lines). Nevertheless, other modes, such as $m=0$ and $m=2$, do remain stable until $t \sim 75 \ta$. After this point, the energy of these modes continues rising with the same growth rate as that of $m=1$.  The kinetic energy modes $m=1$ and $m=2$ evolve in a similar way.
The mode $m=0$ has a larger energy which is roughly constant along the linear phase. 

Fig.~\ref{fig:m_energy_evo_std}(c) corresponds to simulation A\_R150B10, with the rotation rate increased by a factor of two (i.e,  $\eta\sim 14$).  The magnetic mode $m=1$ grows linearly until $t\sim 40 \ta $, yet the same mode of the kinetic energy seem to be stable.  Interestingly, the magnetic and kinetic modes $m=1$ and $2$ (and other high order modes not presented in the figure) at both, poles and equator reach the energy of saturation following two subsequent energy jumps. The same is observed in the evolution of $\tilde{E}_B$ and $\tilde{E}_K$ for simulations with $\eta > 14$, i.e., a slow initial linear growth, with a growth rate that depends on the rotation rate (see Table~\ref{tab:results}), followed by energy jumps.  The panel (d) of Fig.~\ref{fig:m_energy_evo_std}, shows the evolution of simulation A\_R10B10, with $\eta = 197$, where the initial growing is almost entirely suppressed. 

Fig.~\ref{fig:m1_energy_comp_begin} displays a comparison between the initial temporal evolution of the mode magnetic $m=1$ for the simulations of the set A1 Table~\ref{tab:results}). The thicker lines correspond to the simulations presented in Fig.~\ref{fig:m_energy_evo_std}. The thin lines correspond to the remaining simulations as seen in the legend.   
The first $ \sim 15 \ta$ of the non-rotating simulation (A\_NRB10) have been disregarded because they correspond to initial transient period described above. 
Note the decreasing growth rate of the simulations as a function of $\eta$, in this case varied through the rotation rate.  The jumps of energy occur after $ \sim 40 \ta$ for $\eta \gtrsim 13$. 
For intermediate values of $\eta$ the TI coexist with the process generating these energy jumps. 
In the fast rotating simulations, $\eta \gtrsim 100$,  the TI seems to be suppressed but the surges still occur.  These results partially agree with the local simulations performed by \citet{Braithwaite2006StbTor} which focus only in the early stages of the linear phase. Yet, these energy jumps are not observed in the time evolution of their models.  

Fig.~\ref{fig:r10_m1_comparison_wa} shows the temporal evolution of simulations where $\Omega_0$ is kept constant and $\eta$ is changed by considering different values of $B_0$ (set A3).   The results confirm the same pattern discussed above, i.e., for large $\eta$ (panel a) the TI is suppressed and the energy grows through consecutive jumps. For $\eta \lesssim 100$, the TI coexists with these energy surges.  Note that these cases cannot be directly compared with the simulations of set A1, because varying $B_0$ also changes the value of $\delta$. And both quantities influence the growth rate.

In EULAG-MHD the dissipation coefficients depend on the resolution of the model. 
Generally, it is expected that viscosity, thermal conductivity, and magnetic diffusivity 
decrease as resolution increases. As previous studies have demonstrated, changes in these coefficients also affects the TI. 

Finally, Fig.~\ref{fig:m_energy_evo_high} shows the 
time evolution of simulations in set B, where the grid resolution is increased by a factor of two in all directions. Panels (a) to (c) corresponds to different value of $\eta$ obtained with different rotation rates. Line styles and colors are the same as in Fig.~\ref{fig:m_energy_evo_std}. 
The resolution increase implies the reduction of thermal and magnetic diffusion coefficients, which have concurrent effects. A lower thermal diffusion makes the system more stable to TI, while lower magnetic diffusion makes it more unstable.
Panel (a) shows that this did not significantly changed the non-rotating case displayed in Fig.~\ref{fig:m_energy_evo_std}(a). The major differences being the lower growth rate and consequent delayed saturation phase. This decrease in $\sigma$ suggest a dominance of the thermal effects in this case. 
This result agrees with previous analytical \citep[see e.g.][]{Spruit1999} and numerical \citep{Braithwaite2006StbTor} results.
Meanwhile, for rotating simulations the change in $\sigma$ is insignificant. For Panel (b) even the moment of saturation phase is similar to its counterpart in Fig.~\ref{fig:m_energy_evo_std}(b). Meanwhile, the fast rotating case demonstrates a tenfold increase in the stabilization during linear phase.

From all the sets of simulations presented in this section, it is possible to conclude that rotation stabilizes the development of the Tayler instability. Larger values of $\eta$ and $\delta$ are able to fully suppress the development of the $m=1$ mode for a few tens of Alfven travel times. Afterwards, a new instability occurs showing sudden surges of energy. Note that for very slow rotation, e.g., case A\_R300B10, these jumps are not observed. While to this point we are not able to explain the physics of this second instability, the sets of simulations A1, A3 and B, show that its development depends on the magnetic field strength and the 
dissipation coefficients. 
The stronger the field, and the smaller the 
dissipation coefficients, the longer it takes for the jumps to occur. Independently of the nature of these energy surges, in all the cases the final energy of the unstable modes never surpasses the initial magnetic energy. Since the model solves for the full system of non-linear equations, it is possible to explore the long term behavior of the magnetic field in the simulations. The next sections presents the results concerning to the subsequent saturation and decaying stages.

\subsubsection{Saturation phase}

\begin{figure}
        \includegraphics[width=\columnwidth]{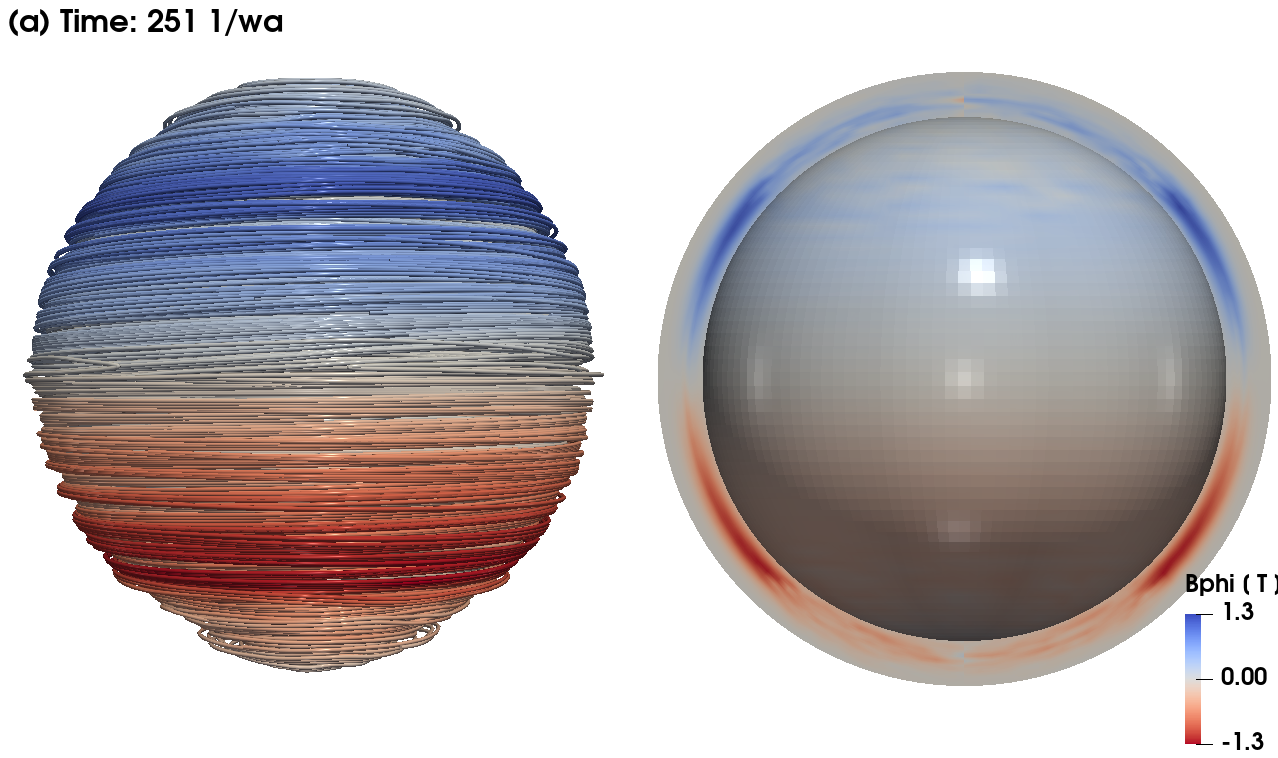}
        \includegraphics[width=\columnwidth]{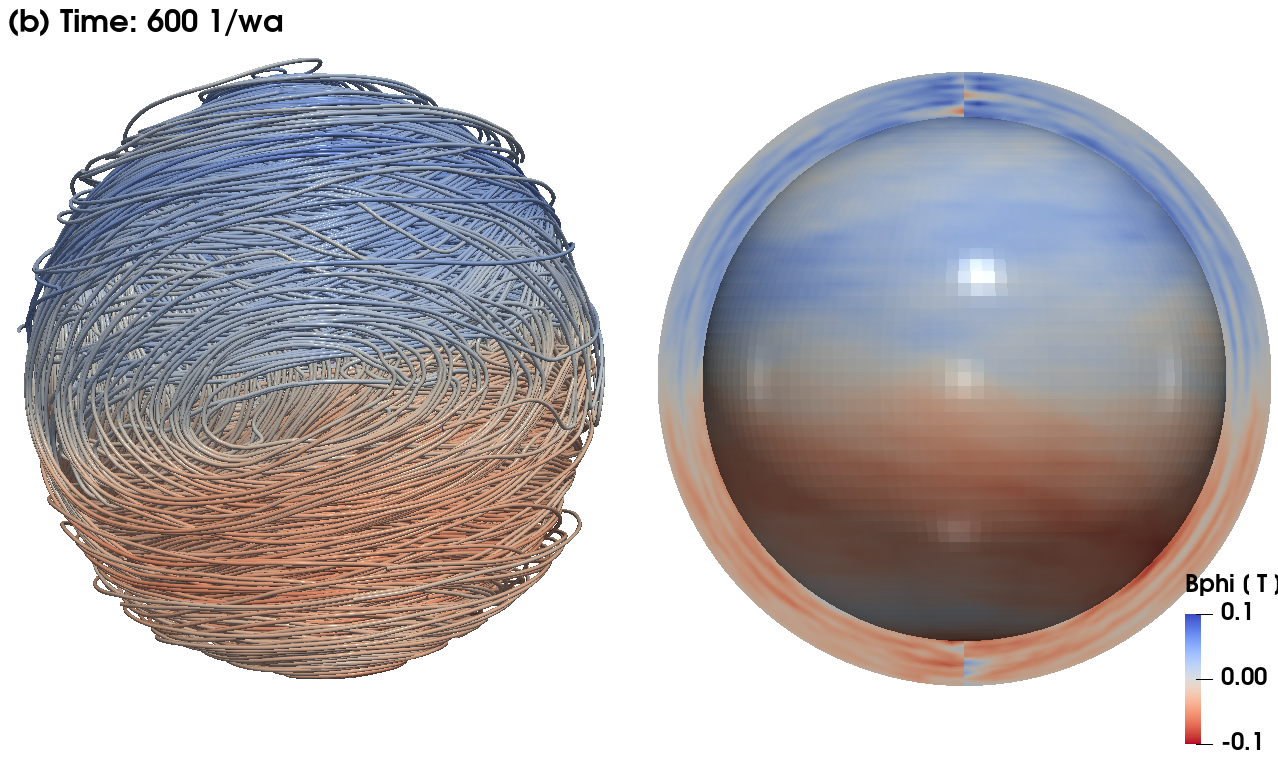}
\caption{Similar to Fig.\ref{fig:initial_B} for simulation A\_R300B10 at (a) end of linear phase ($251 \: 1/\omega_A$) and (b) during the decaying phase ($600 \: 1/\omega_A$).}
\label{fig:A_r300_snap}
\end{figure}

\begin{figure}
  \includegraphics[width=\columnwidth]{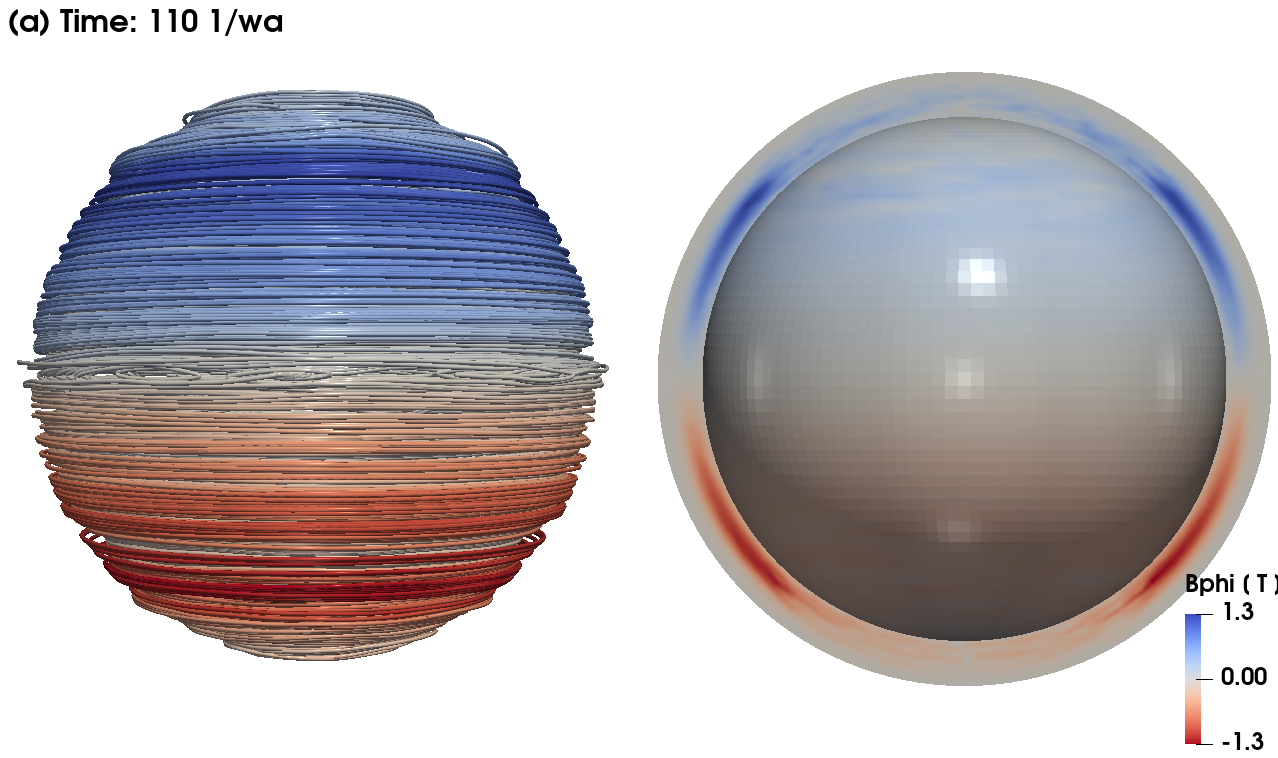}
  \includegraphics[width=\columnwidth]{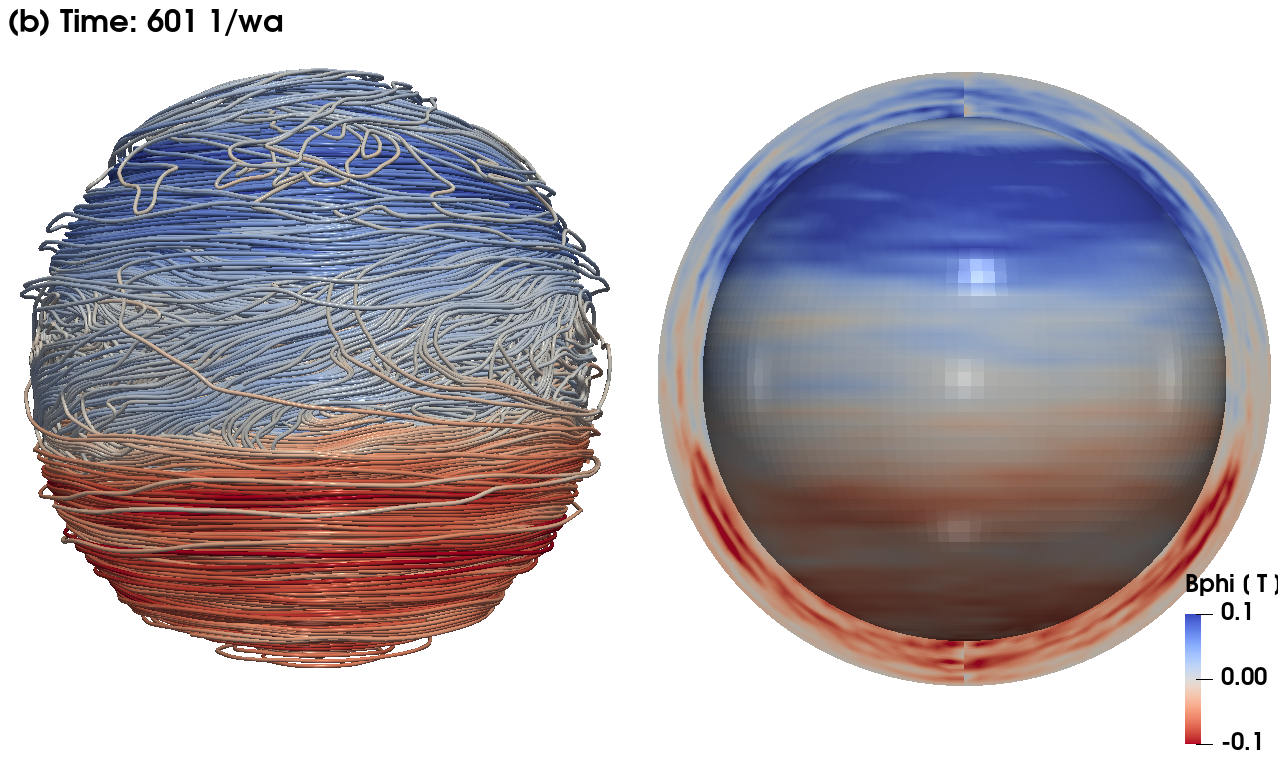}
\caption{Similar to Fig.~\ref{fig:initial_B} for simulation A\_R10B10 at (a) end of linear phase ($110 \: 1/\omega_A$) and (b) during the decaying phase ($601 \: 1/\omega_A$).}
\label{fig:Anti_10r_10b_snap}
\end{figure}

\begin{figure*}
  \subfloat[Non-rotating]{\label{subfig:field_lines_nr}
	\minipage{0.31\textwidth}
	\begin{center}
	   \includegraphics[width=\linewidth]{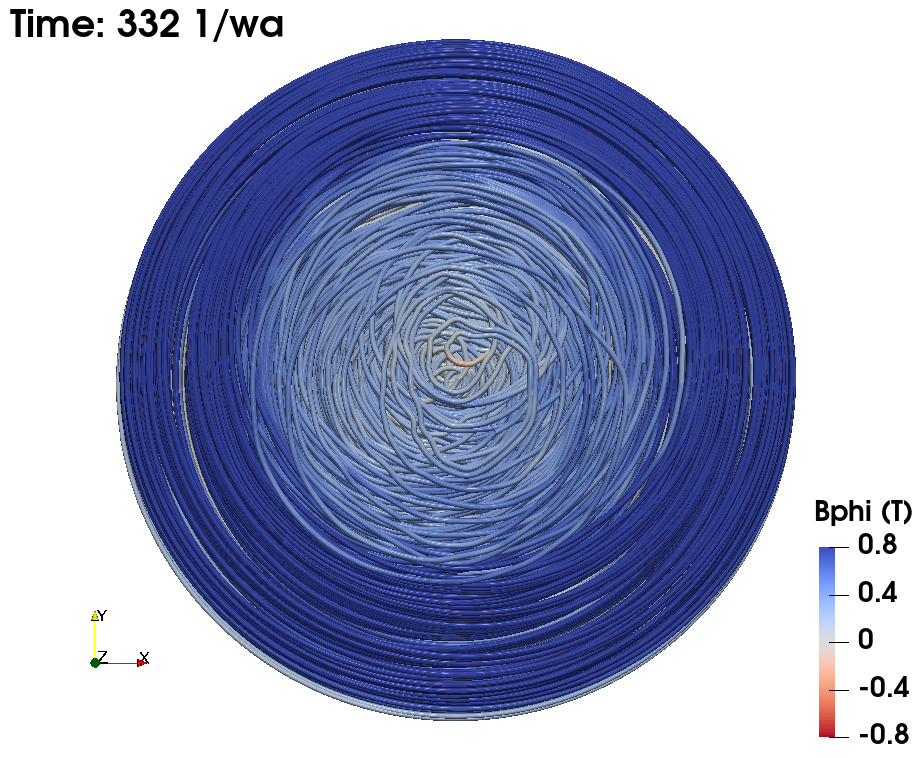}
	   \includegraphics[width=\linewidth]{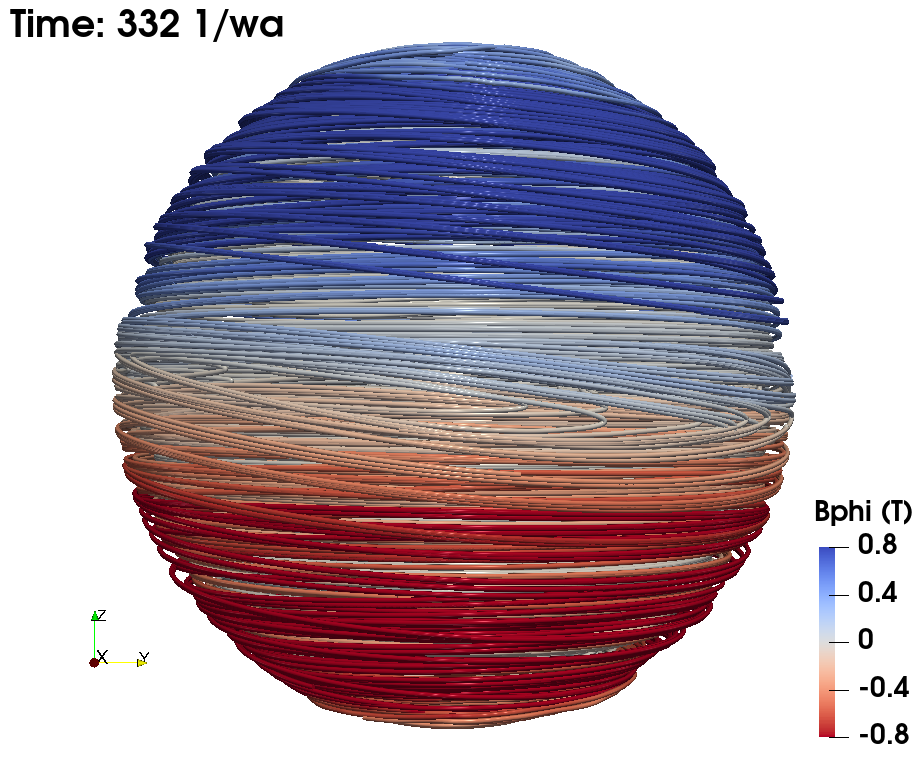}
	\end{center}
	\endminipage}
  \subfloat[Slow rotation (300 days)]{\label{subfig:field_lines_r300}
	\minipage{0.31\textwidth}
	\begin{center}
	   \includegraphics[width=\linewidth]{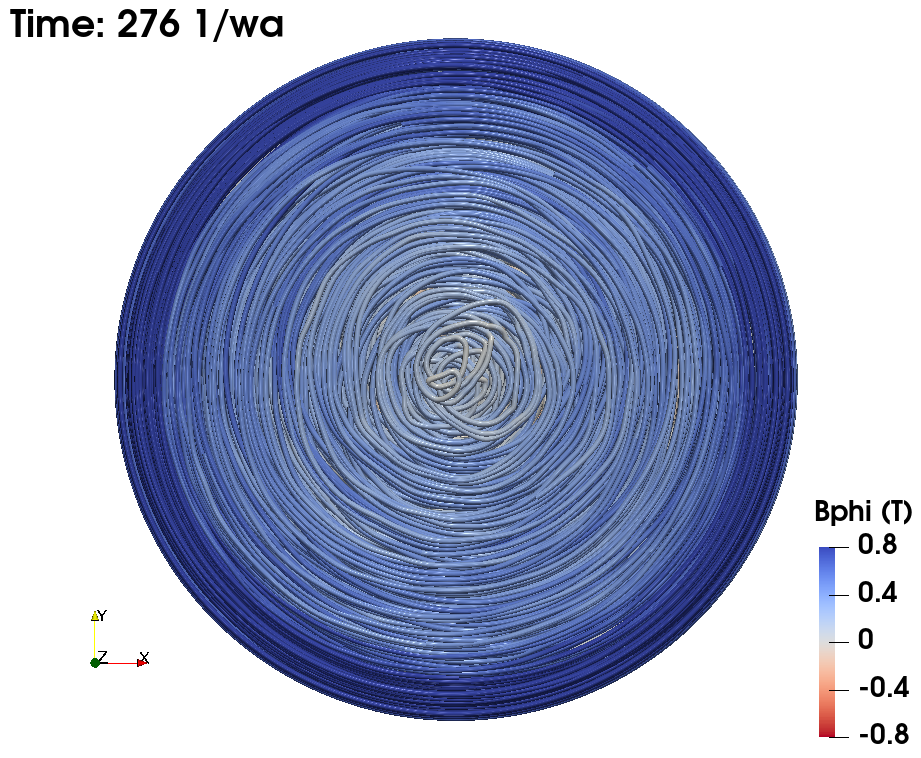}
	   \includegraphics[width=\linewidth]{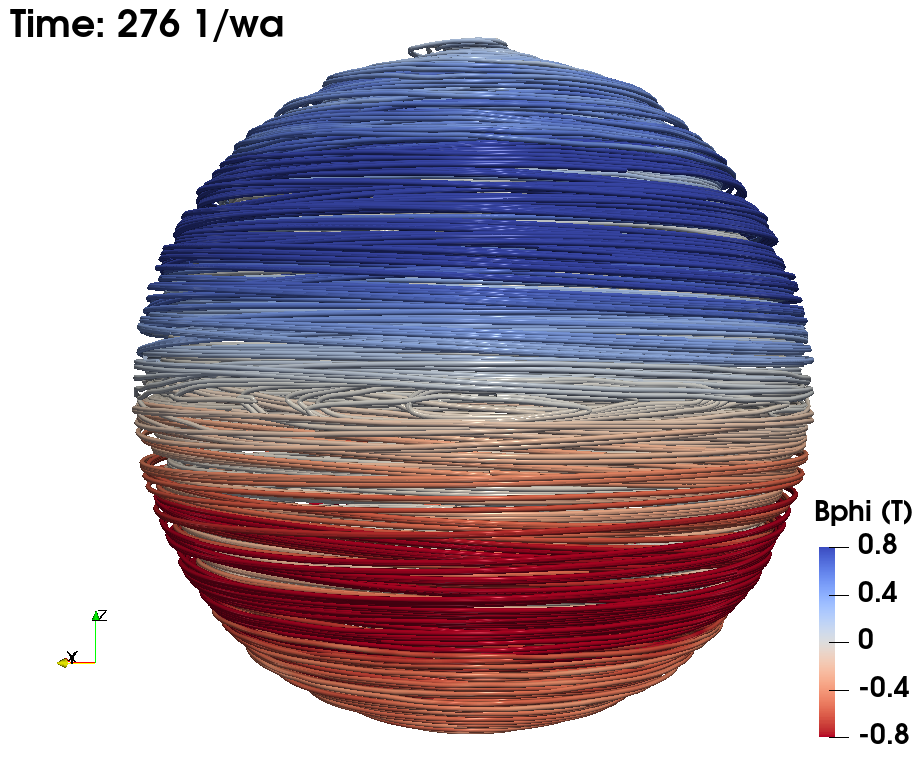}
	\end{center}
    	\endminipage}
  \subfloat[Fast rotation (10 days)]{\label{subfig:field_lines_r10}
	\minipage{0.31\textwidth}
	\begin{center}
	   \includegraphics[width=\linewidth]{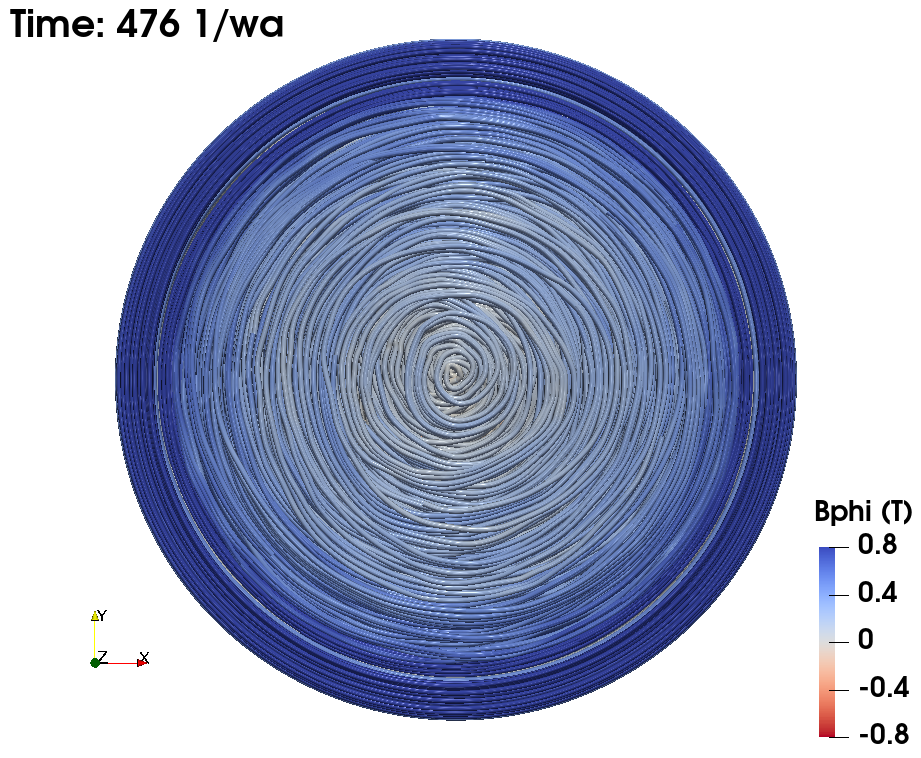}
	   \includegraphics[width=\linewidth]{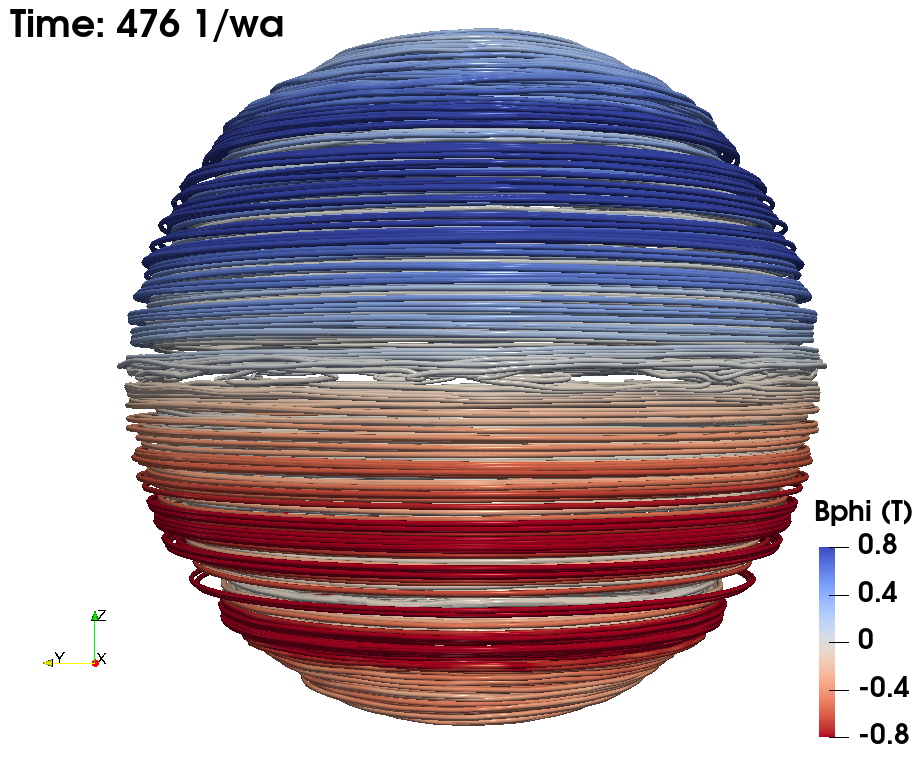}
	\end{center}
	\endminipage}
\caption{
Magnetic field lines of the (a) non-rotating (B\_NRB10), (b) slow rotating (B\_R300B10), and (c) fast rotating (B\_R10B10) simulations of set B. The snapshots are taken at the time of energy saturation of the unstable modes,  and at $r = 0.68 \Rsun$. The top panels show a projection of the geometrical north pole, the bottom panels show a equatorial projection. The color scheme is the same as in Fig.~\ref{fig:initial_B}. Animations showing the temporal evolution of the magnetic field lines for these simulations are available as supplementary material. }
\label{fig:field_lines}
\end{figure*}

Once the unstable modes reach saturation energy, a change in the topology of the initial magnetic field is observed. The saturation, decay, as well as the final configuration of the magnetic field depend on the rotation rate.  Figs.~\ref{fig:A_r300_snap} and \ref{fig:Anti_10r_10b_snap} show the magnetic field lines from the simulation with the slowest (A\_R300B10, $\eta=6.7$) and fastest rotation (A\_R10B10, $\eta=197$), respectively.  On each figure the panels (a) depict the saturation stage where the topology still contains a prominent toroidal field, yet the displacement of the field lines show the imprint of the unstable modes. In the panels (b), the snapshots are taken during the decaying phase. The magnetic field strength is about one order of magnitude smaller than its initial strength, and the field lines are deformed mostly at equatorial latitudes. However, a toroidal structure is still observed. This is in contrast with the results obtained for the non-rotating case (A\_NRB10, $\eta=0$), presented in  Fig.~\ref{fig:Anti_nr_10b_snap}, and demonstrates the stabilizing effects of rotation. Also, note that the magnetic field contains a relevant poloidal component (vertical field lines) which may contribute to preserve a stable configuration.

The same behavior in the saturation stage is observed in the high resolution simulations (set B). A comparison between cases B\_NRB10 ($\eta=0$), B\_R300B10 ($\eta=6.7$)  and B\_R10B10 ($\eta=197$) 
is presented in Fig.~\ref{fig:field_lines}(a)-(c), respectively.   The color coding in this figure is the same as in Fig.~\ref{fig:initial_B}. The upper (lower) row shows projections of  the north pole (equator).  In panels (a) and (b), the mode  $m=1$ appears at the poles as a misaligned structure that wobbles around the rotation axis. At the equator, it appears as the opening of the field lines forming a {\it clamshell}-like structure.  Note also the large inclination of the magnetic axis resulting in the non-rotating case. On the other hand,  the field lines obtained in the fast rotating simulation (panel c), do not display relevant tilt of the magnetic axis,  and the field lines at the equator are barely open, depicting only high order modes.

\begin{figure}
  \includegraphics[width=\linewidth]{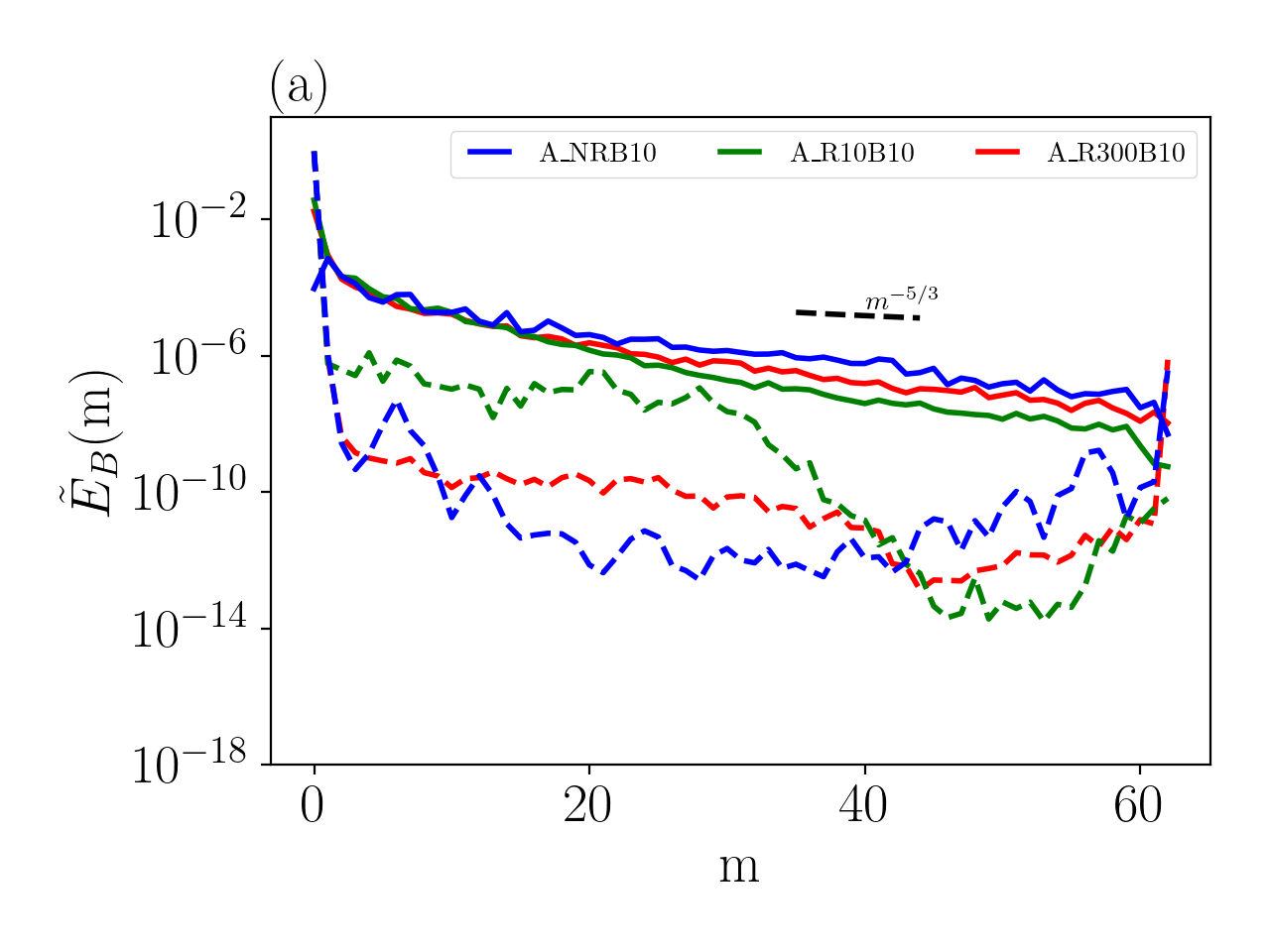}\\
  \includegraphics[width=\linewidth]{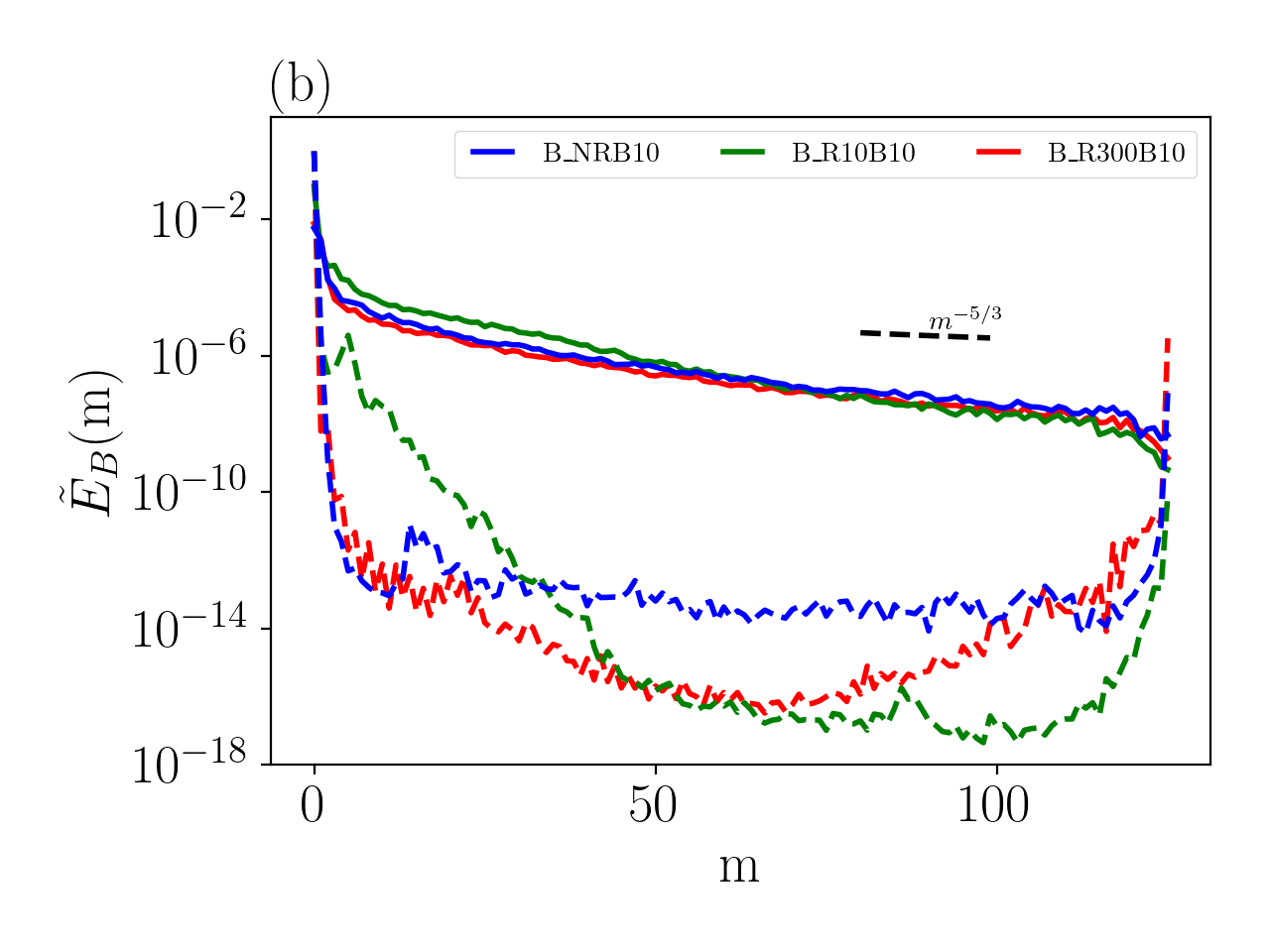}

  \caption{Distribution of the spectral magnetic energy density, $\hat{E}_B$, in the longitudinal modes, $m$,  for simulations (a) of set A, A\_NRB10 (blue line), A\_R10B10 (green), and A\_R300B10 (red);  and (b) of set B, B\_NRB10 (blue line), B\_R10B10 (green), and B\_R300B10 (red).  The dashed lines correspond to the linear phase in a stage with similar energy between simulations. The continuous lines correspond to the decaying phase. 
  In this phase we observe the slope typical of fully developed homogeneous and isotropic turbulence. Animations showing the temporal evolution of the magnetic and kinetic energy spectra, for simulations $A\_R10B10$ and $B\_R10B10$, are available as supplementary material.
  }
  \label{fig:m_modes_energies}
\end{figure}

The qualitative description of the results in Figs.~\ref{fig:Anti_10r_10b_snap} and \ref{fig:field_lines} can be quantified through the spectrum of the magnetic energy density. In Fig.~\ref{fig:m_modes_energies} the energy spectrum, {$\hat{E}_B$}, is calculated by decomposing the total vector magnetic field on its spherical harmonics representation using the SHTns library \citep{SHTNS2013}, at the radius $r=r_c$. Note that in this case the spectrum averages all the  
available latitudinal modes.  The upper and lower panels correspond to characteristic simulations from set A1, A\_NRB10 (blue line), A\_R300B10 (red), A\_R10B10 (green); and from set B, B\_NRB10 (blue),  B\_R300B10 (red), and B\_R10B10 (green), respectively.  The dashed (continuous) lines corresponds to the linear (decaying) phase at an evolution time where the energy in the mode $m=1$ is similar in all the presented simulations.  Animations showing the temporal evolution of the spectra are presented in the supplementary material.

During the linear phase (dashed lines in Fig.~\ref{fig:m_modes_energies}), the non-rotating (blue) or slow rotating (red) simulations behave similarly.  The magnetic energy is mostly in the modes $m=0$ (the initial magnetic field) and $1$ (the fastest growing mode). Conversely,  in the simulations with fast rotation (green) modes up to $m \sim 35$ develop similar energy than the mode $m=1$.  These results are independent of the grid resolution.  The energy growth of the high order modes occurs during the energy surges discussed above. At the decaying phase,  the profile of the energy density spectra is similar for all the simulations.   They have a scaling law compatible with decaying turbulence, where $\hat{E}_B \propto m^{-5/3}$ \citep{Kolmogorov1941}. Note that the energy of the $m=0$ mode decays substantially only for the non-rotating simulations (see continuous blue lines). This implies that even slow rotation is sufficient to conserve the initial magnetic field. 


\begin{figure}
  \includegraphics[width=\linewidth]{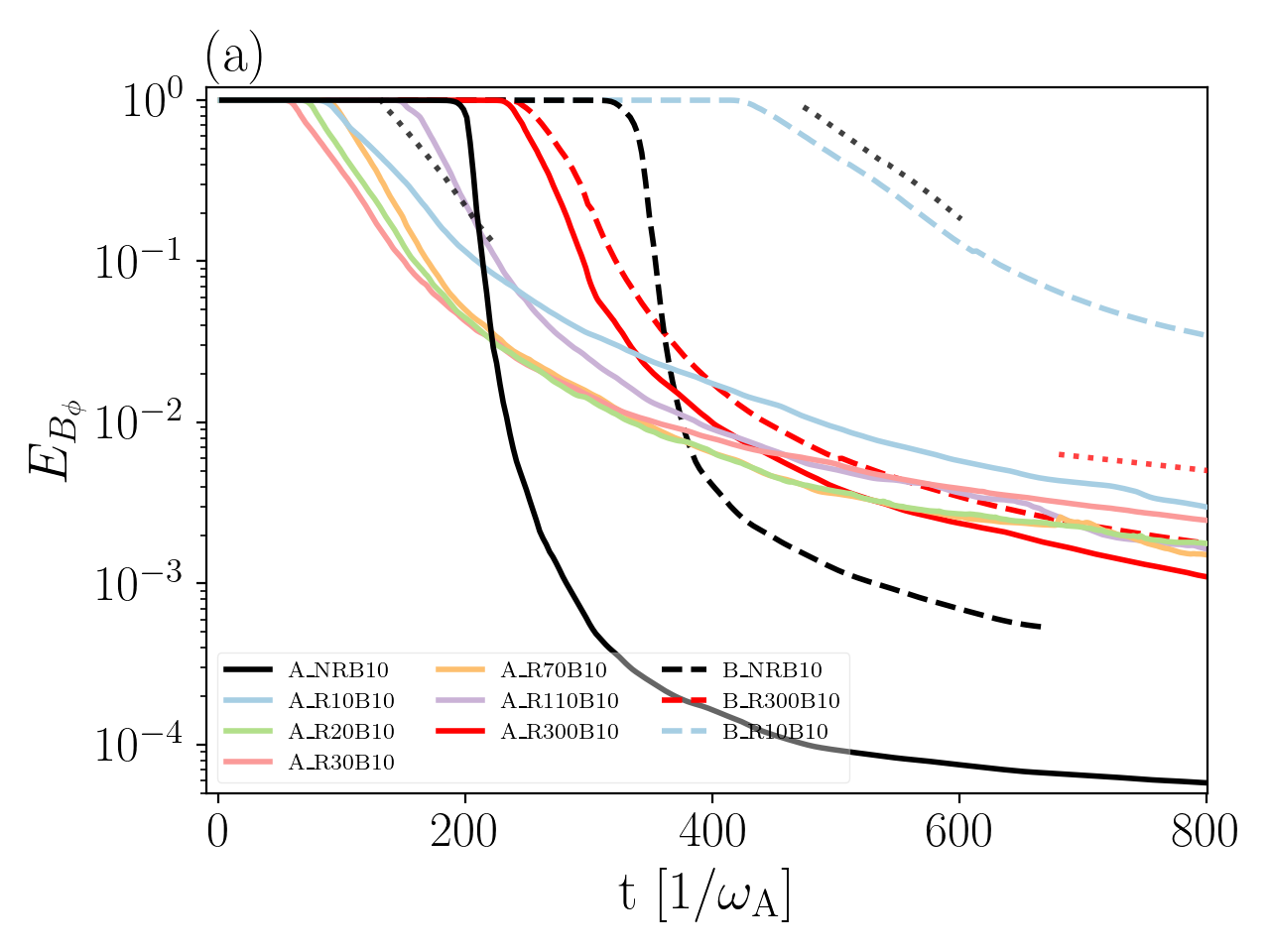}\\
  \includegraphics[width=\linewidth]{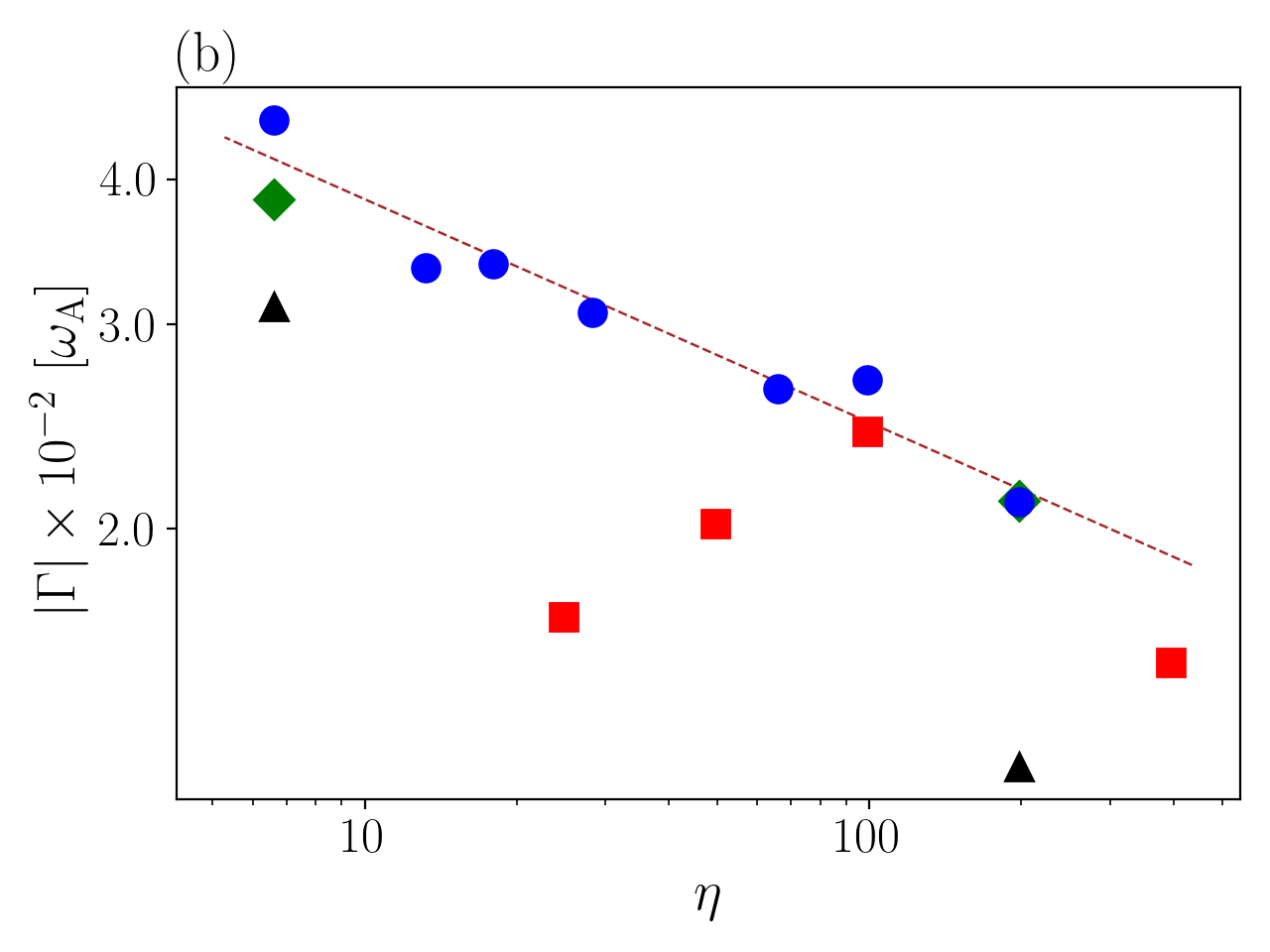}
  \caption{(a) Evolution of $E_{{B}_\phi}$, as a function of time. The continuous (dashed) lines correspond to simulations from the set A1 (B). Lines with different colors correspond to various rotation rates as indicated in the legend. The black dotted lines show the time segments where the decay rate, $\Gamma$, is calculated (for the sake of clarity of the figure, only the segments corresponding to three characteristic simulations are shown). The turbulent magnetic diffusivity, $\eta_{\rm t}$, is estimated in the late phase of evolution ($ t \ge 500 $ [1/$\omega_{\rm A}$]) where the energy decays roughly at the same rate for all simulations (see red dotted line as an example).
  (b) Distribution of $\Gamma$ as a function of $\eta$ (Eq.~\ref{eq:deltaeta}) for simulations 
  from the sets A1 (blue circles), B (black triangles), C (green diamonds).   The simulations from set A2, with fixed rotation and varying magnetic fields, are presented with red squares 
  (see Table~\ref{tab:results}). The inset presents the same results but with $\Gamma$ expressed in terms of the Alfven travel time. }
  \label{fig:m_mode_ts}
\end{figure}

The effects of rotation during the non-linear phase of evolution can be explored through the temporal evolution of the energy contained in the toroidal component of the magnetic field,  $E_{B_{\phi}} = \overline{B^2}_\phi/B_0^2$. Here, the average in $\overline{B^2}$ is taken in the directions $\phi$,  $\theta$, and $r$ between $0.65 \Rsun$ and $0.70 \Rsun$,  corresponding to the full width at half maximum (FWHM) of the initial Gaussian profile of Eq.~(\ref{eq:antisym_magfield_ini}). Fig.~\ref{fig:m_mode_ts}(a) depicts the temporal evolution of $E_{B_{\phi}}$, the continuous (dashed) lines  correspond to simulations from the set A1 (B). Lines with different colors correspond to various rotation rates. The figure encompasses the linear, saturation, and the decaying phases until $t \sim 800 \ta$. Note that the decay of  $E_{B_{\phi}}$ during the linear phase is not evident in this representation because it is of the order of $10^{-6}$ (energy of the residual toroidal field).  The graph demonstrates that rotation prevents the decay of the toroidal magnetic field.  For no rotation the decay is sharp (see black lines). Progressively increasing the rotation rate results in slower and slower decay. 
The decay rate of the toroidal magnetic energy, $\Gamma$, can be estimated in time segments where the time derivative of $E_{B_{\phi}}$ is constant (see the values of $\Gamma$ in Table~\ref{tab:results}). The black dotted lines are eye guides exemplifying these segments.

In addition, $\Gamma$ also depends on the numerical resolution, i.e., on the effective numerical magnetic diffusion. This is evident by comparing the continuous with the dashed lines corresponding to the sets of simulations A1 and B, respectively.  As a matter of fact, increasing the resolution only in the radial coordinate (set C, not presented in the figure), does not show the significant change in the decay rate observed when the resolution is increased twofold in the three coordinates (set B). This results confirms that the evolution of the field happens mainly in the horizontal directions whenever $\delta$ is sufficiently high.

In panel (b) of Fig.~\ref{fig:m_mode_ts} we present $|\Gamma|$ as a function of $\eta$ for simulations from the sets A1 (blue circles), B (black triangles), C (green diamonds). The value of $|\Gamma|$ changes by a factor of 2.15 from the lowest to the highest rotating rates, also lowest to highest values of $\eta$ (simulations A\_R300B10 to A\_R10B10, blue circles), with a scaling that may be fitted with the power law $|\Gamma| \propto \eta^{-0.19}$.  Extrapolating these law for $|\Gamma| \rightarrow 0$ (fully stable situation) implies unrealistic fast rotation. Nevertheless, the black triangles, corresponding to the set B, show that this result is sensitive to the resolution. For instance, the decay rate changes by a factor of 1.8 between simulations A\_R10B10 and B\_R10B10 ($\eta = 197.8$), and of 1.7 between simulations A\_R300B10 and B\_R300B10 ($\eta = 6.7$). 

The red squares show the changes in $|\Gamma|$ for simulations with a fixed rotation rate, $\Omega_0 = 7.27 \cdot 10^{-6}$ Hz, and variations in $\eta$ due to different amplitudes of $B_0$ (see simulations A\_R10B5 to A\_R10B80 in Table~\ref{tab:results}).  
These cases show a curve with low values of $|\Gamma|$ for the smaller values of $\eta$ (large magnetic fields), a maximum at $\eta=98.9$, and a decay for the larger values of $\eta$ (weaker magnetic fields).  This trend happens because changing the values of $\eta$ in this form, also changes the normalization time scale.  In physical units, for $B_0$ greater (smaller) than $1.3$ T,  the values of $|\Gamma|$ are higher (lower) than those corresponding to the set A1.  
Thus, it is possible to conclude that the decay of the initial magnetic field is less affected by the TI whenever the relevant time scale is set by the rotation.

\subsection{Long term evolution}

After the sharp decay observed during the saturation phase, $E_{B_{\phi}}$ continues decreasing.  Interestingly, 
during this phase the slope of the curves changes and ultimately tends to have a similar slope, independent of the rotation rate or even of the numerical resolution (see $t > 700 \ta$ in Fig.~\ref{fig:m_mode_ts}(a)). This may be interpreted as a turbulent diffusive decay also confirmed by the $m^{-5/3}$ scaling observed in the magnetic power spectra (the same scaling is obtained for the kinetic power spectra, not presented in the paper).  Therefore, by multiplying the $|\Gamma|$ estimated for this stage (see red dotted line if Fig.~\ref{fig:m_mode_ts}(a)) by $\varpi_c^2/2$, it is possible to evaluate the turbulent magnetic diffusivity, $\beta_{\rm t}$.  For the set A1 of simulations $\beta_{\rm t} \sim 8 \cdot 10^{7}$  $\mathrm{m^2/s}$; for set B, $\beta_{\rm t} \sim 5\cdot10^{7}$ $\mathrm{m^2/s}$.  This may have consequences regarding the life time of magnetic fields in SSLs.     

\begin{figure}
\includegraphics[width=1\columnwidth]{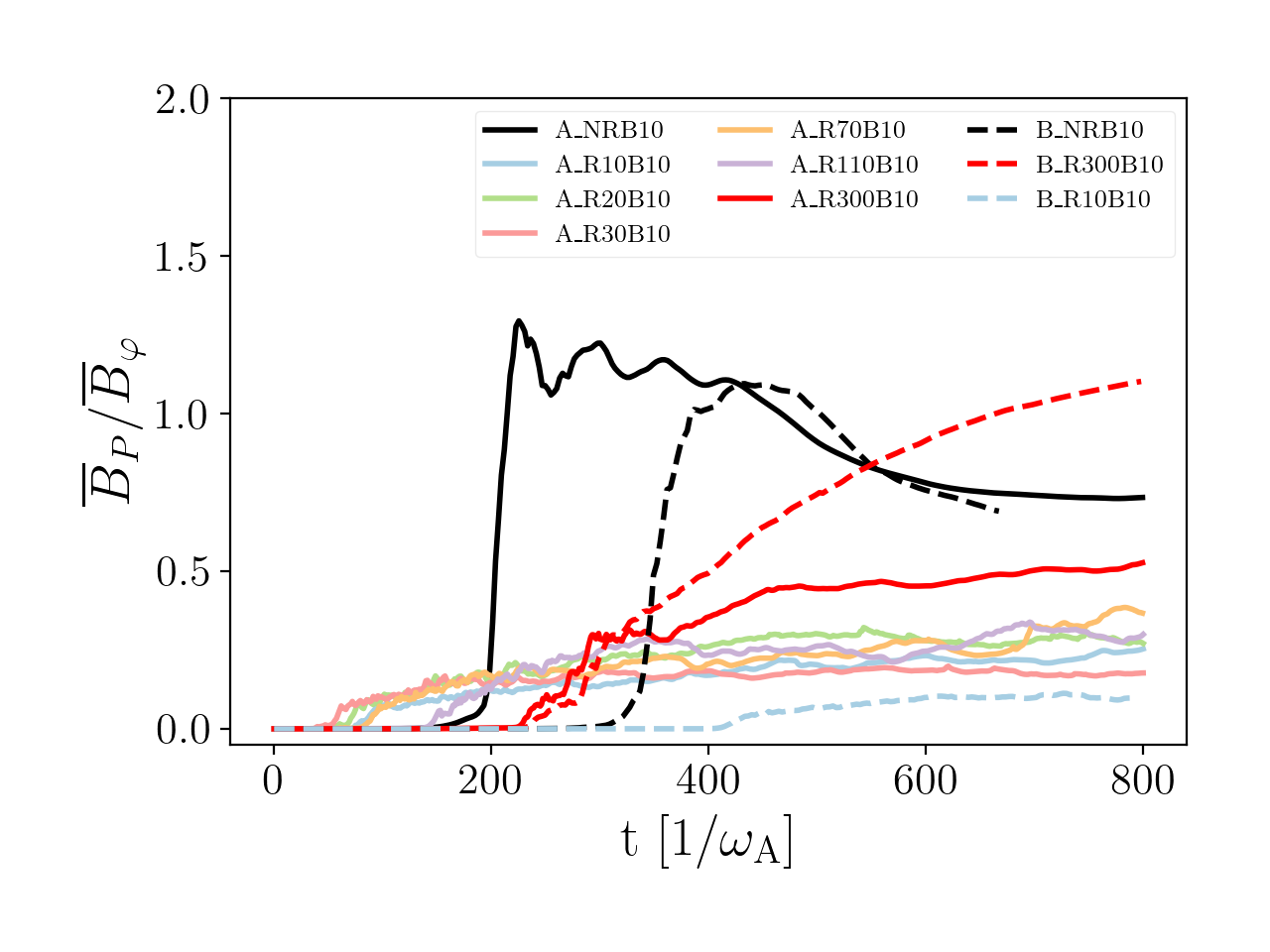}
\caption{Temporal evolution of the ratio $\overline{B}_P/\overline{B}_{\phi}$ for simulations in
the sets A (continuous lines) and B (dashed lines).  Note that rapidly rotating simulations seem 
to converge to a constant value of this ratio.}
\label{fig:BpBt_comp}
\end{figure}

The decay of toroidal field observed in Fig.~\ref{fig:m_mode_ts}(a) is  followed by an increase of the poloidal component of the magnetic field which ultimately leads to a stable configuration.   Figure~\ref{fig:BpBt_comp} displays the ratio between the poloidal and the toroidal field components, $\overline{B}_P/\overline{B}_{\phi}$, in terms of the Alfven travel time. Here,
$\overline{B}_P = \sqrt{\overline{B_\theta^2} + \overline{B_r^2}}$ and $\overline{B}_\phi$ are the volume averaged poloidal and toroidal magnetic field components, respectively. The average is performed as described above.  
Continuous and dashed lines correspond to simulations of sets A1 and B, respectively. The black lines represent the non-rotating cases, while different colors correspond to different rotation rates.

A significant increase in the poloidal field is observed when the simulations reach the saturation phase (Fig.~\ref{fig:BpBt_comp}). The non-rotating cases, as well as the cases with slow rotation, $\eta=6.7$,
display a rapid increase of the poloidal component, followed by a decrease when the simulation advances towards the decaying phase. 
For simulations with $\eta \gtrsim 30$, the ratio $\overline{B}_P/\overline{B}_{\phi}$  converges to a plateau at
$\sim 0.3$. The same ratio is also observed for the set of simulations A3 where both parameters $\delta$ and $\eta$ change.  In conclusion, 
both, the rotation and the poloidal field contribute to sustain the remnant toroidal field stable for several hundreds of Alfven travel times. Nevertheless, this field is embedded in a turbulent region and decays accordingly. 
For the simulation with higher resolution B\_R10B10, the final value of $\overline{B}_P/\overline{B}_{\varphi}$ is smaller than for their low resolution counterpart, A\_R10B10 (see dashed light-blue line in Fig.~\ref{fig:BpBt_comp}). Therefore, although the initial field is unstable, the growth of the unstable modes is prevented by rotation. Furthermore, only a small fraction of poloidal field leads to a configuration that remains stable, yet, this field is prone to turbulent diffusion.   With the increase of resolution, the saturation and decay stages occur at a much longer time (compare solid and dashed light blue lines in Fig.~\ref{fig:m_mode_ts}(a)). This makes higher resolution simulations highly time consuming and prohibitive for the available supercomputer time. Thus, it remains uncertain how the ratio $\overline{B}_P/\overline{B}_{\varphi}$  will change for effective diffusion approaching the values of the Ohmic magnetic diffusivity inside radiative zones. For the radiative zone below the solar tachocline, $\beta_{\rm Ohm} \sim 10^3$ m$^2$/s \citep{Zahn2007}. 

\subsection{Helicities}

\begin{figure}

  \includegraphics[width=\columnwidth]{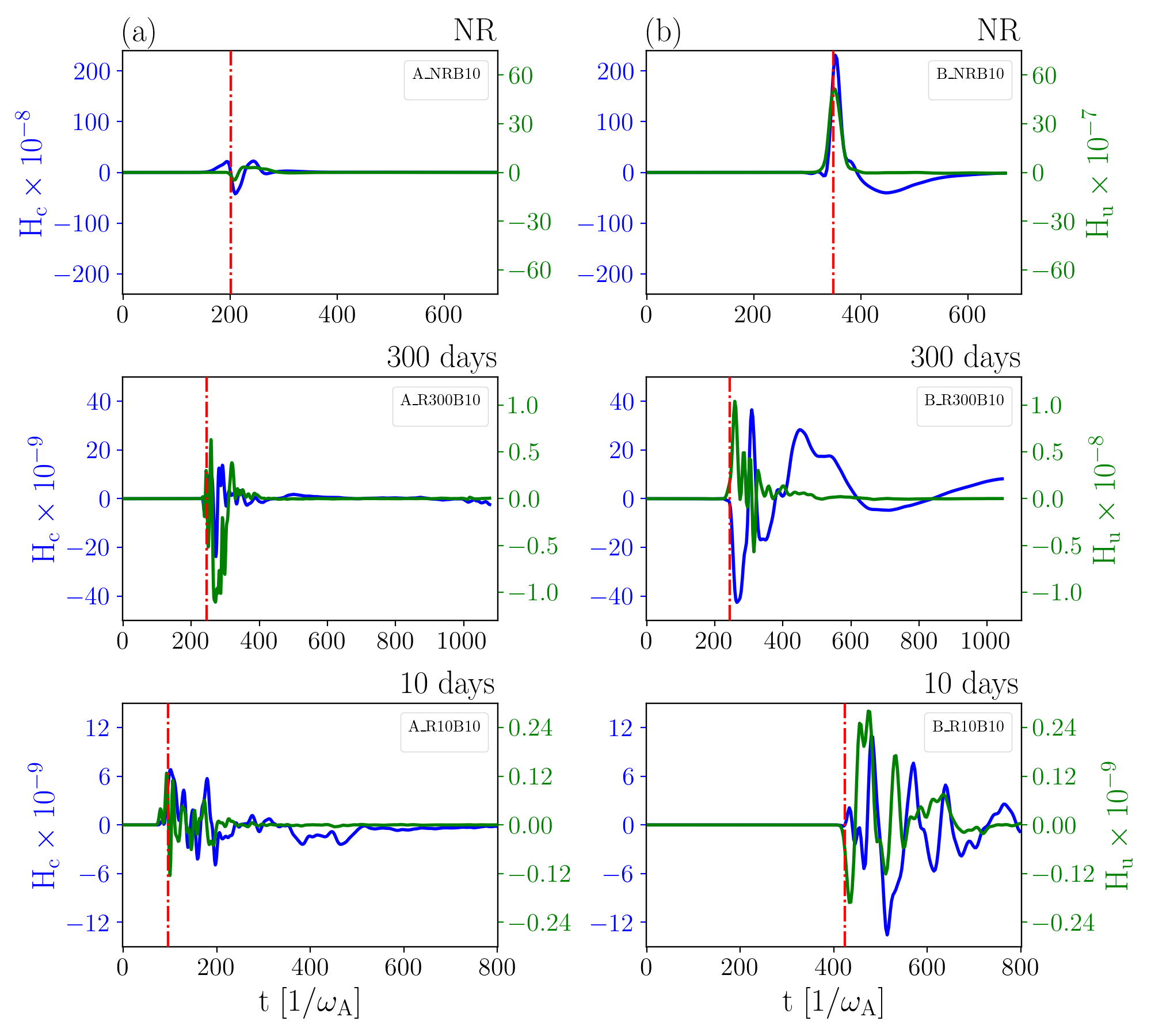}
  \caption{ Temporal evolution of $H_c$ (blue) and $H_u$ (green) for simulations in sets A (column a) and B (column b), with each row corresponding to the same rotational period. The intensities of the helicities were multiplied by a factor of $5$ on column a for better visualization.
  }
  \label{fig:helicity}
\end{figure}

The generation of the poloidal magnetic field observed in Fig.~\ref{fig:BpBt_comp} must originate from the processes discussed in the previous sections. This field component develops as the magnetic field attempts to achieve a stable configuration \citep{Tayler1973}.  As seen, the rotating cases reach a steady state with a magnetic field mostly toroidal with a small fraction of the poloidal component. On the other hand, the poloidal field amplification is associated with the development of helical motions or currents in dynamo theory \citep{SKR66}. Note, however, that the fastest growing mode of the magnetic field associated to the Tayler instability is $m=1$,  whereas in the mean-field dynamo theory the large scale is normally considered axisymmetric. Therefore, mean-field dynamo coefficients require some axisymmetric contribution.  By computing the small-scale kinetic and current helicities from the simulated data we can infer if such contribution exists.

In Fig.~\ref{fig:helicity} we depict the time evolution of both the small scale kinetic and current helicities,
\begin{eqnarray}
H_u =&\overline{(\boldsymbol{\nabla}\times\boldsymbol{u}^\prime)\cdot\boldsymbol{u}^\prime}\\\nonumber
H_c =&  {\overline{(\boldsymbol{\nabla}\times\boldsymbol{B}^\prime)\cdot\boldsymbol{B}^\prime}}/(\mu_0 \rho_{ad}) ,   
\end{eqnarray}

where the prime denotes perturbations of the velocity and magnetic fields with respect to the longitudinal mean, i.e., all modes but $m=0$. The overline here has the same meaning previously defined, except that now the latitudinal average is done in the Northern hemisphere only. In Fig.~\ref{fig:helicity}, the green and blue lines corresponds to the
$H_u$ and $H_c$, respectively. The left and right columns show three characteristic simulations from set A1, namely A\_NRB10, A\_R300B10 and A\_R10B10, and three from the set B, B\_NRB10, B\_R300B10 and B\_R10B10.

Signatures of the helicities are observed for all cases, starting to appear just before the saturation phase (see red vertical line). This time correlates well with the increase of $\overline{B}_P/\overline{B}_{\phi}$ observed in Fig.~\ref{fig:BpBt_comp}.  Nevertheless, there is no preferred sign for the helicities, but sign reversals occur in all cases.  
The amplitudes of $H_u$ and $H_c$ have the same order of magnitude, yet the signal of  $H_u$  decays faster.  It is clear, however, that the helicities from simulations of set B have higher amplitudes.  Notably, the non-rotating cases (first row) display a stronger current helicity than the rotating counterparts. Meanwhile, in the rotating simulations the helicities last for several Alfven travel times, depicting several sign reversals and a damped evolution.  

Previous studies have demonstrated that large scale shear in combination with a sign changing may lead to incoherent $\alpha$-effect and may be sufficient for sustaining the dynamo \citep[e.g.][]{VB1997ApJ,MB2012MNRAS}. 
Thus, the results presented here might support the idea of a large-scale dynamo in a radiative layer. Simulations including shear are ongoing work. 

\section{Summary and conclusions}
\label{sec:conclusion}
Magnetic fields are presumably present in the radiative zone of solar-like stars and have also been observed at the surface of massive Ap/Bp objects. However, certain magnetic fields topologies in these layers are prone to different kinds of instabilities \citep[e.g.,][]{Tayler1973,MarkeyTayler1973}.

This study focuses on the Tayler instability (TI), where the decay of an initial axisymmetric toroidal field leads to the growing of other longitudinal modes, modifying the original configuration. \citet{Tayler1973, PittsTayler1985} proposed that this instability could be quenched by the the influence of rotation and/or by the presence of a poloidal field  whose forces opposes the displacement caused by the unstable modes. In this work, we use non-linear MHD simulations, in spherical geometry,  to explore the stabilizing role of rotation on a toroidal magnetic field, anti-symmetric about the equator, permeating a layer whose 
values of the Brunt-Väisälä frequency are similar to those of the lower part of the solar tachocline.
The simulations were performed with the EULAG-MHD code, which solves the anelastic MHD equations. To understand the role of rotation in the Tayler instability we consider non-rotating and rotating cases, with rotational periods between 300 and 10 days.  Alternatively, we present cases with fixed rotation and various initial magnetic field strengths. Our analysis was performed in terms of the non dimensional quantities, $\eta$  (Eq.~\ref{eq:deltaeta}), which measures the relative importance of the Coriolis and magnetic forces, and $\delta$, measuring the relative importance of the buoyant and magnetic forces.  

Since the magnetic diffusivity,  as well as other dissipative processes, are expected to be insignificant in the solar radiative zone, we use the code without explicit dissipation terms. Nevertheless, there is always a numerical effective dissipation of all quantities which in the EULAG-MHD code is nonlinear and intermittent in space and time, and depends on the grid size. Thus, increasing the numerical resolution of the model results in less dissipation. This affects the dynamics of the magnetic field in different forms. We remark that simultaneously diminishing thermal diffusion and magnetic diffusivity may have opposite effects regarding the TI \citep{Zahn1974,Braithwaite2006StbTor}. 

During the evolution of the magnetic field we identify three stages. First, the linear phase, where the unstable modes grow exponentially. Most of the previous work regarding TI has focused on this phase. Second, when the growing modes reach their maximum energy, the initial magnetic field exhibits a sharp decay. This is identified as the saturation phase. And third, a diffusive decay phase. At this stage the energy spectrum follows a scaling law with $E \sim m^{-5/3}$. Thus, we identify this as turbulent decay.

For all sets of simulations presented in Table~\ref{tab:results}, we find that the cases without rotation or rotating slowly ($\eta < 6.7$)  present a clear exponential growth of the longitudinal mode $m=1$ during the linear phase. This is the canonical signature of the Taylor instability.  Although the unstable modes develop first at polar regions, there is no significant difference in the growth rate between polar and equatorial latitudes.  

For $\eta > 13.4$, the influence of rotation appears and the mode $m=1$ is kept stable for several Alfven travel times. However, after this stable phase, sudden energy surges occur for the modes $m \ge 1$,  which subsequently reach energy levels close to the energy of $B_{\phi 0}$.  This is an unexpected result and, to our knowledge, not reported in the literature.  We have been unable to identify the nature of these energy surges and they will be subject of follow up studies.
Nevertheless,  the results from simulations in sets A3 and B indicate that the time interval where the modes $m \ge 1$ remain stable depends on the amplitude of the initial magnetic field and on the resolution of the simulations, tantamount of the dissipative processes.  These dependencies make us believe that this is a physical phenomenon rather than a numerical artifact. 

Either because of the TI or the energy surges, the unstable modes reach energy levels compatible to that of the initial field, $B_{\phi 0}$, i.e., in both cases the energy of the growing modes is extracted from the initial magnetic field which decays at this saturation stage.  Consequently, a substantial change in the magnetic field topology is observed during the saturation phase. The decay rate of these changes, $|\Gamma|$, depends on the value of $\eta$. The results demonstrate that the initial magnetic field topology is less affected by the TI whenever the relevant time scale is set by the rotation.

As the toroidal field decays, the formation of a poloidal components is observed in all simulations. The analysis shows that for simulations with $\eta \gtrsim 30$, the ratio $\overline{B}_P/\overline{B}_{\phi}$ converges to a plateau at $\sim 0.3$.  The same ratio is also observed for the set of simulations A3 where both parameters, $\delta$ and $\eta$, change. For the high resolution simulation B\_R10B10, this ratio is smaller,   $\overline{B}_P/\overline{B}_{\phi} \sim 0.1$. The implication of this finding is that, despite the initial instability, the growing of unstable modes is effectively prevented by the Coriolis force.  These modes transfer only a small fraction from the initial magnetic energy into a poloidal field. The resultant configuration remains topologically stable by hundreds of Alfven travel times (see Fig.~\ref{fig:field_lines}). 

After reaching such a ratio, the simulations evolve with this equilibrium configuration prone only to turbulent diffusion.  For these cases, despite the decay in amplitude, the field experiences minor topological changes. On the other hand, in cases without rotation or with lower $\eta$, the initial topology is mostly lost. The poloidal component grows rapidly and eventually leads the simulation to stop, requiring prohibitive smaller time steps. 

During the generation of poloidal field, prior to the saturation phase, the simulations develop helical motions and currents. They do not show a preferred hemispheric sign, but several reversals with damping evolution.  Interestingly, the amplitude of the helicities is larger for the non-rotating simulations. However, in these cases the helicities also decay faster.  Such incoherent $\alpha$-effect may be sufficient for sustaining a dynamo. Thus, the results presented here might support the idea of a large-scale Tayler-Spruit dynamo in a radiative layer.  To sustain this dynamo, however, a certain amount of shear is required to replenish the toroidal field.  The evolution of the field under these conditions is left for a future study.

\section*{Data availability statement}
Analysis results and data underlying this article will be shared on request to the corresponding author.

\section*{Acknowledgment}

GM would like to thank the support from CAPES. GG acknowledges
COFFIES Phase II NASA grant 80NSSC22M0162. FDS acknowledges support from a Marie Curie Action of the European Union (Grant agreement 101030103). National Center for Atmospheric Research (NCAR) is sponsored by the National Science Foundation. The work has been performed under the Project HPC-EUROPA3 (INFRAIA-2016-1-730897), with the support of the EC Research Innovation Action under the H2020 Programme; in particular, the authors gratefully acknowledges the support of INAF OACT and the computer resources and technical support provided by Cineca.



\bibliographystyle{mnras}
\bibliography{references.bib} 






\bsp	
\label{lastpage}
\end{document}